\documentclass[a4paper,11pt]{article}
\pdfoutput=1 
\usepackage{jheppub}

\usepackage[T1]{fontenc} 
\usepackage[applemac]{inputenc}

\usepackage{amsmath, amssymb, dsfont}
\usepackage[numbers]{natbib}
\usepackage{graphicx}
\usepackage{enumerate}
\usepackage[normalem]{ulem}
\usepackage{empheq}
\usepackage[justification=centerlast, singlelinecheck=false, caption=false]{subfig}

\newcommand{\rd}{\mathrm{d}}
\newcommand{\re}{\mathrm{e}}

\newcommand{\comp}[2]{(\!\begin{smallmatrix}#1\\#2\end{smallmatrix}\!)} 
\newcommand{\Mp}{M_\mathrm{Pl}}
\newcommand{\N}{\mathcal{N}}
\newcommand{\C}{\mathcal{C}}

\title{\boldmath SLED Phenomenology: Curvature vs.\ Volume}

\author[a,c]{Florian Niedermann}
\author[a,c]{and  Robert Schneider}

\affiliation[a]{Arnold Sommerfeld Center for Theoretical Physics, Ludwig-Maximilians-Universit\"at, Theresienstra{\ss}e 37, 80333 Munich, Germany}
\affiliation[c]{Excellence Cluster Universe, Boltzmannstra{\ss}e 2, 85748 Garching, Germany\\
~\\}
 
\emailAdd{florian.niedermann@physik.lmu.de}
\emailAdd{robert.bob.schneider@physik.lmu.de}

\abstract{
We assess the question whether the SLED (Supersymmetric Large Extra Dimensions) model admits phenomenologically viable solutions with 4D maximal symmetry. We take into account a finite brane width and a scale invariance (SI) breaking dilaton-brane coupling, both of which should be included in a realistic setup. Provided that the microscopic size of the brane is not tuned much smaller than the fundamental bulk Planck length, we find that either the 4D curvature or the size of the extra dimensions is unacceptably large. Since this result is independent of the dilaton-brane couplings, it provides the biggest challenge to the SLED program.

In addition, to clarify its potential with respect to the cosmological constant problem, we infer the amount of tuning on model parameters required to obtain a sufficiently small 4D curvature. A first answer was recently given in~\cite{Niedermann:2015via}, showing that 4D flat solutions are only ensured in the SI case by imposing a tuning relation, even if a brane-localized flux is included. In this companion paper, we find that the tuning can in fact be avoided for certain SI breaking brane-dilaton couplings, but only at the price of worsening the phenomenological problem.

Our results are obtained by solving the full coupled Einstein-dilaton system in a completely consistent way. The brane width is implemented using a well-known ring regularization. In passing, we note that for the couplings considered here the results of~\cite{Niedermann:2015via} (which only treated infinitely thin branes) are all consistently recovered in the thin brane limit, and how this can be reconciled with the concerns about their correctness, recently brought up in~\cite{Burgess:2015kda}.
}

\begin{document}

\maketitle
\flushbottom


\section{Introduction and Summary}

The SLED model~\cite{Aghababaie:2003wz} provides a promising candidate for addressing the cosmological constant (CC) problem~\cite{Weinberg:1988cp}. The main motivation is that for a codimension-two brane, the 4D CC only curves the transverse extra-space into a cone, while the on-brane geometry stays flat. However, it was realized from the very beginning~\cite{Aghababaie:2003wz} that for compact extra dimensions this comes at the price of yet another tuning relation, stemming from the flux quantization condition, which in turn is required to stabilize the compact extra space. Alternatively, from a 4D point of view, the problem can be formulated as saying that it is simply the classical scale invariance (SI) of this theory which leads to a flat brane geometry, in which case Weinberg's general no-go argument~\cite{Weinberg:1988cp} applies.

To circumvent this problem, a brane-localized flux (BLF) term was later included~\cite{Burgess:2011va, Burgess:2011mt}; the idea was that if this term breaks SI, then it is in principle possible that the dilaton dynamically adjusts such that flux quantization is fulfilled, thereby avoiding the tuning relation (or runaway solutions). However, it was recently shown~\cite{Niedermann:2015via} (and also confirmed in a specific UV model~\cite{Burgess:2015gba}) that only SI brane couplings---including the BLF term---ensure a flat brane geometry. But then, it does not alter the tuning (or runaway) problem either, and we are basically back at square one.

However, the mere fact that the 4D curvature is zero in the SI case does not immediately rule out the model as a potential solution to the CC problem. It might still be possible to achieve a \emph{nonzero but small} (compared to standard model loop contributions) curvature in a phenomenologically viable and technically natural way by breaking SI on the brane. The main purpose of this companion paper to~\cite{Niedermann:2015via} is to investigate this remaining question in detail.

The starting point of our analysis is the effective theory that is obtained after solving for the Maxwell field in a 4D maximally symmetric configuration, and adding a counter-term to dispose of divergences which generically arise due to the BLF, as discussed in~\cite{Niedermann:2015via}. The goal here is to explicitly solve the resulting Einstein-dilaton system for given model parameters and couplings.
Explicitly, we will focus on a SI breaking brane tension. Since the standard model sector breaks SI on the brane, this term should be included in a realistic setup, and its size will be set by loop contributions of the brane matter fields. Furthermore, we will endow the brane with a finite thickness in order to avoid potential divergences. This should not be viewed as a mere technical regularization, but rather as another physically unavoidable feature: A realistic brane has to come with some microscopic thickness, which would ultimately be determined by an underlying UV model. We will find that both sources---the non SI tension and the brane width---contribute to the 4D curvature independently, and discuss them in detail.

To endow the brane with a thickness, we choose in Sec.~\ref{sec:ring_reg} a convenient and well-known technique~\cite[see e.g.][]{Peloso:2006cq, Burgess:2008yx} that replaces the infinitely thin brane by a ring of finite proper circumference $ \ell $. Most importantly, we expect the low energy questions we are going to ask to be insensitive to this microscopic choice. This setup only admits static solutions if there is some additional mechanism that prevents the ring from collapsing. Effectively, this boils down to adding an angular pressure component $ p_\theta $, the size of which can be inferred from the junction conditions across the brane. This allows us to generalize a previously derived formula for the 4D curvature to the regularized setup, thereby enabling us to study the tuning issue and the phenomenological viability of the model. Prior to that, we check in Sec.~\ref{sec:delta_limit} whether our result are consistent with the delta-analysis in~\cite{Niedermann:2015via}:  We find that the delta-results are all recovered in the thin brane limit if and only if $ p_\theta \to 0 $. Since for an infinitely thin object there is no direction this pressure could act in, this is a reasonable physical assumption.\footnote{Nonetheless, it was recently disputed in~\cite{Burgess:2015kda} and used as an argument against the trustworthiness of~\cite{Niedermann:2015via}. We comment on this in Appendix~\ref{ap:Cliff}.} Here, it will also be shown to be true for the case of exponential dilaton-brane couplings as introduced in Sec.~\ref{sec:near_SI}. These couplings model the SI breaking and are of particular interest with respect to the CC problem as they allow to be close to SI without the need of tuning the coefficients small.

A discussion of the model's phenomenological status is given in Sec.~\ref{sec:phen}, leading to an unambiguous conclusion: {\it Without tuning certain model parameters to be small compared to the bulk Planck scale, it is not possible to comply with both the observed value of the Hubble parameter as well as constraints on the size of the extra dimensions.} This negative conclusion applies to both the SI breaking tension and the finite brane width effects independently.

This---so far analytical---verdict is based on several assumptions that are all confirmed by explicitly solving the brane-bulk system in Sec.~\ref{sec:num_results}. To that end, the full set of field equations for a 4D maximally symmetric ansatz is integrated numerically, as explained in Sec.~\ref{sec:num_alg}. Special attention is given to imposing the required regularity conditions at \emph{both} axes of the compact space, because only then are all integration constants uniquely determined. The results and physical implications, both for SI and non SI dilaton-brane couplings, are discussed in Secs.~\ref{sec:SI} and~\ref{sec:non_SI}, respectively. 
We find that in both cases an acceptably small 4D curvature is typically only achieved by tuning the (dilaton independent part of the) brane tension, but that this tuning can indeed be alleviated for certain brane-dilaton couplings. However, we also confirm the analytic prediction, so that in either case the extra dimensions are way too large to be phenomenologically viable.

Let us note that the same model was recently analyzed in~\cite{Burgess:2015lda} in a dimensionally reduced, effective 4D theory. Our present work instead solves the full 6D bulk-brane field equations, thus providing an alternative and complementary approach. While confirming the result of~\cite{Burgess:2015lda} that a large extra space volume can be achieved for certain parameters without the need for putting in large hierarchies by hand, we are also able to go one step further and uncover the tuning that is always needed to get \emph{both} the 4D curvature \emph{and} the volume within their observational bounds.
Our conclusions are summarized in Sec.~\ref{sec:concl}.

\section{Delta Brane Setup}
\label{sec:thin_brane}

\subsection{Review}

We first provide a brief review of the thin brane setup. The reader familiar with the corresponding discussion in our companion paper \cite{Niedermann:2015via} should feel free to skip this section.

The field content of the SLED model comprises the 6D metric $g_{AB}$, a Maxwell field $A_{B}$, which stabilizes the compact bulk dimensions, and the dilaton $\phi$, which renders the bulk theory SI. The corresponding action reads~\cite{Burgess:2011mt}
\begin{equation} \label{eq:action}
S = S_\mathrm{bulk} + S_\mathrm{branes} \,,
\end{equation}
where the bulk part is\footnote{We use the same notation and conventions as in~\cite{Niedermann:2015via}.}
\begin{equation} \label{eq:action_bulk}
	S_\mathrm{bulk} = - \int \rd^6 X \sqrt{-g}\,\left\{\frac{1}{2\kappa^2}\left[ R + (\partial_M\phi)(\partial^M\phi) \right] + \frac{1}{4} \re^{-\phi}F_{MN} F^{MN} + \frac{2e^2}{\kappa^4}\re^{\phi}\right\} ,
\end{equation}
with  $\kappa$ and $e$ the gravitational and U(1) coupling constants, respectively. The 6D Ricci scalar $R$ is built from the 6D metric $g_{AB}$, and $ F \equiv \rd A $. The brane contributions are
\begin{equation} \label{eq:action_brane}
S_\mathrm{branes} = - \sum_b \int \rd^4 x \sqrt{-g_4} \left\{ \mathcal{T}_b(\phi)-\frac{1}{2} \mathcal{A}_b(\phi) \epsilon_{mn} F^{mn} \right\} ,
\end{equation}
where the index $b \in \{+,-\}$ runs over both branes situated at the north ($+$) and south ($-$) pole of the compact space, where the metric function $ B $ (see below) vanishes. The 4D brane tension is denoted by $\mathcal{T}_b(\phi)$. The second term, controlled by $\mathcal{A}_b(\phi)$, describes the brane localized flux (BLF). In general, both terms are allowed to have arbitrary dilaton dependences; in particular, the SI case corresponds to $\mathcal{T}_b(\phi)=\mathrm{const}$ and $\mathcal{A}_b(\phi) \propto \re^{-\phi}$.

In \cite{Niedermann:2015via} we investigated  the theory under the assumption of 4D maximal symmetry and azimuthal symmetry in the bulk. This leads to the following general ansatz,
\begin{subequations} \label{eq:ansatz}
	\begin{align}
	\rd s^2 & = W^2(\rho) \,\hat g_{\mu\nu} \rd x^{\mu} \rd x^\nu + \rd \rho^2 + B^2(\rho) \rd \theta^2 \,, \label{eq:ansatz_met}\\
	A  &= A_\theta(\rho) \rd\theta \,, \\
	\phi & = \phi(\rho) \,, \label{eq:ansatz_phi}
	\end{align}
\end{subequations}
where $\hat g_{\mu\nu}$ is 4D maximally symmetric and thus fully characterized by its (constant) 4D Ricci scalar $ \hat R $.
With these symmetries, the Maxwell equations can be integrated analytically, yielding
\begin{equation} \label{eq:sol_F}
	F_{\rho\theta} = \re^{\phi} B \left[\frac{Q}{W^4} + \sum_b \frac{\delta_b}{2\pi B} \mathcal{A}_b(\phi) \right] ,
\end{equation}
where $ Q $ is an integration constant, and $ \delta_b $ is shorthand for the Dirac delta function $ \delta(\rho - \rho_b) $. In the case of a nonvanishing BLF, the second term leads to a divergence $\propto \delta(0)$ in the remaining equations of motion, which can be interpreted as a relict of treating the branes as point-like objects. We proposed a corresponding brane counter term which allowed to consistently dispose of this contribution. After this subtraction, the remaining field equations consist of the dilaton equation
\begin{equation}
\label{eq:dilaton}
	-\frac{1}{\kappa^2}\frac{1}{BW^4}\left( B W^4 \phi' \right)' = \frac{\re^{\phi}}{2}\left( \frac{Q^2}{W^8} - \frac{4e^2}{\kappa^4} \right)
	- \sum_b \frac{\delta_b}{2\pi B} \left\{ \mathcal{T}^{\prime}_b(\phi) - \frac{Q}{W^4} \re^{\phi} \left[ \mathcal{A}^{\prime}_b(\phi) + \mathcal{A}_b(\phi) \right] \right\} \,,
\end{equation}
and the $ \comp{\mu}{\nu} $, $ \comp{\rho}{\rho} $ and $ \comp{\theta}{\theta} $ components of Einstein's field equations,
\begin{subequations} \label{eq:einstein_expl}
	\begin{align}
	-\frac{1}{\kappa^2} \left( \frac{\hat R}{4 W^2} + 3 \frac{W''}{W} + \frac{B''}{B} + 3\frac{W'^2}{W^2} + 3 \frac{W'B'}{WB} + \frac{1}{2} \phi'^2 \right) 
	& = \frac{\re^\phi}{2}  \left( \frac{Q^2}{W^8} + \frac{4e^2}{\kappa^4} \right) \nonumber\\
	& \quad + \sum_b \frac{\delta_b}{2\pi B} \mathcal{T}_b(\phi)
	\,, \label{eq:einstein_00}
	\\
	\frac{1}{\kappa^2} \left( \frac{\hat{R}}{2W^2} + 6\frac{W'^2}{W^2} + 4\frac{W'B'}{WB} - \frac{1}{2} \phi'^2 \right) 
	& = \frac{\re^{\phi}}{2} \left( \frac{Q^2}{W^8} - \frac{4e^2}{\kappa^4} \right) \,, \label{eq:einstein_rho}
	\\
	\frac{1}{\kappa^2} \left( \frac{\hat{R}}{2W^2} + 4\frac{W''}{W} + 6\frac{W'^2}{W^2} + \frac{1}{2} \phi'^2 \right)
	& = \frac{\re^{\phi}}{2} \left( \frac{Q^2}{W^8} - \frac{4e^2}{\kappa^4} \right) \,. \label{eq:einstein_theta}
	\end{align}
\end{subequations}
%

Integrating the dilaton equation over an infinitesimally small disc covering one of the axes yields the boundary condition for $\phi$. For  $W$ and $B$ the same is achieved by taking appropriate combinations of the Einstein equations. Explicitly, one finds
\begin{subequations}
	\label{eq:matching_delta}
	\begin{align}
	\left[B \phi' \right]_{\rho=\rho_b} &= \frac{\kappa^2}{2\pi} \C_b \,, \label{eq:jump_delta_phi}\\
	\left[B (W^4)' \right]_{\rho=\rho_b} &=0 \,,\\
	[B']_{\rho=\rho_b} &= 1 - \frac{\kappa^2}{2\pi} \left [ \mathcal{T}_b(\phi) \right ]_{\rho=\rho_b} \,,
	\end{align}
\end{subequations}
where we defined
\begin{equation} \label{def:SI_combination}
	\C_b := \left\{ \mathcal{T}^{\prime}_b(\phi) - \frac{Q}{W^4} \re^{\phi} \left[\mathcal{A}^{\prime}_b(\phi) + \mathcal{A}_b(\phi)\right] \right\}_{\rho=\rho_b} \,,
\end{equation}
which measures the brane coupling's deviation from SI. 

Furthermore, integrating a suitable combination of the field equations over the whole compact extra space yields
\begin{equation} \label{eq:degrav_cond_delta}
	V \hat R  = 2 \kappa^2 \sum_b  W_b^4 \C_b \,,
\end{equation}
with the 2D volume defined as
\begin{equation}
	\label{def:volume}
	V := 2\pi \int \!\rd\rho\, BW^2 = \int \rd^2 y \, \sqrt{g_2} \, W^2 \,.
\end{equation}
Hence, the SI case ($ \mathcal{C}_b = 0 $) implies $ \hat{R} = 0 $.

\subsection{Constraint}\label{sec:constraint}

Let us now turn to a peculiarity~\cite{Burgess:2015kda} of the delta setup which was not discussed in~\cite{Niedermann:2015via}. Multiplying the constraint~\eqref{eq:einstein_rho} by $ B^2 $ and taking the limit $ \rho \to \rho_b $ yields (assuming that $ B^2 \re^\phi \to  0 $)
\begin{equation} \label{eq:brane_constraint}
	\left\{ \frac{3}{8W^8} \left[ B (W^4)' \right]^2 + \frac{1}{W^4} \left[ B (W^4)' \right] \left[ B' \right] - \frac{1}{2} \left[ B \phi' \right]^2 \right\}_{\rho = \rho_b} = 0 \,.
\end{equation}
The terms in square brackets are those which appear in the boundary conditions~\eqref{eq:matching_delta}, and so we are lead to (assuming that $ \left [ \mathcal{T}_b(\phi) \right ]_{\rho=\rho_b} $ is finite, as it should be for physically relevant situations)
\begin{equation}
	\C_b = 0 \,. \label{eq:SI_delta}
\end{equation}
This is in clear contradiction to the SI breaking expectation $ \C_b \neq 0 $. In~\cite{Burgess:2015kda}, it was argued that this uncovers an inconsistency of the delta analysis; we will comment on this in more detail in Appendix~\ref{ap:Cliff}.
Here, let us merely state the other possibility: that~\eqref{eq:SI_delta} is in fact another prediction of the delta setup, saying that it is impossible to consistently break SI on a delta-brane, at least on-shell. In this work, we will explicitly verify that this option is indeed realized for a relevant class of couplings. More specifically, starting with exponential SI breaking couplings of the form $ \C_b \propto \re^{\gamma\phi_b} $ and a thick brane setup, we will find that $ \phi_b \to -\infty $ in the thin brane limit, thereby restoring $ \C_b \to 0 $.

At this point, let us also emphasize that the SI case is completely insensitive to this whole issue, because then~\eqref{eq:SI_delta} is identically fulfilled. Thus, the important achievement of~\cite{Niedermann:2015via}, namely the first correct identification of those BLF couplings which unambiguously lead to $ \hat{R} = 0 $ (and the resulting tuning relation), remains unaffected.\footnote{In fact, the whole analysis of~\cite{Niedermann:2015via} could also be trivially adapted to the point of view of~\cite{Burgess:2015kda} on the SI breaking case (by simply including an angular pressure $ p_\theta $), without changing any of the conclusions. It would only add another contribution $ \propto p_\theta $ to~\eqref{eq:degrav_cond_delta}, which also only vanishes in the SI case. However, we regard an angular pressure for an infinitely thin object as unphysical, cf.~Sec.~\ref{sec:delta_limit} and Appendix~\ref{ap:Cliff}.}

However, \eqref{eq:SI_delta} also implies that the actual (nonzero) value of $ \hat{R} $ for broken SI cannot be inferred within the pure delta framework (which always\footnote{In the proposal of~\cite{Burgess:2015kda} $ \hat{R} \neq 0 $ would still be possible for delta branes, but only at the price of allowing $ p_\theta \neq 0 $.} predicts $ \hat{R} = 0 $), but requires studying a thick brane setup. This also has the advantage that potential singularities are regularized.


\section{Thick Brane Setup}
\label{sec:thick_brane}


\subsection{Ring Regularization}
\label{sec:ring_reg}

In order to avoid any singularities and potential ambiguities of the (non SI) delta brane setup, the authors in~\cite{Burgess:2015nka, Burgess:2015gba} introduced a specific UV model describing the brane as a vortex of finite width in extra space. We will instead use a different and technically simpler way of regularizing the system, in which the delta brane is replaced by a ring of circumference $\ell$ \cite{Peloso:2006cq, Burgess:2008yx}.\footnote{Note that even though it is not obvious how the BLF term could be consistently adapted to the 5D brane in a covariant way at the level of the action, introducing the regularization after the Maxwell field has been solved for is straightforward. (In any case, the BLF term will in the end not be crucial for our main conclusions.)} We assume the microscopic details of the regularization to be irrelevant for the low energy questions we want to study.

Let us note that introducing the regularization scale $ \ell $ breaks SI. This, however, does not necessarily imply that the underlying UV theory (which would resolve the brane microscopically) breaks SI explicitly. Indeed, a SI mechanism could easily be built, in analogy to the flux stabilization which fixes the large size of the extra dimensions. In that case, the UV model parameters would not determine $ \ell $, but rather the SI combination $ \ell \re^{\phi_0/2} $.\footnote{This is analogous to the SI GGP solutions~\cite{Gibbons:2003di, Niedermann:2015via}, where not the extra space volume $ V $ is fixed, but only the combination $ V \re^{\phi_0} $.} However, this does not change the fact that $ \ell $ has to take a specific value in order to comply with observations. For a SI UV model, this would correspond to a spontaneously broken SI; but the physical conclusions would be the same.



\begin{figure}[t]
	\centering
	\includegraphics[width=0.6\textwidth,trim={2cm 3.5cm 2cm 5cm},clip]{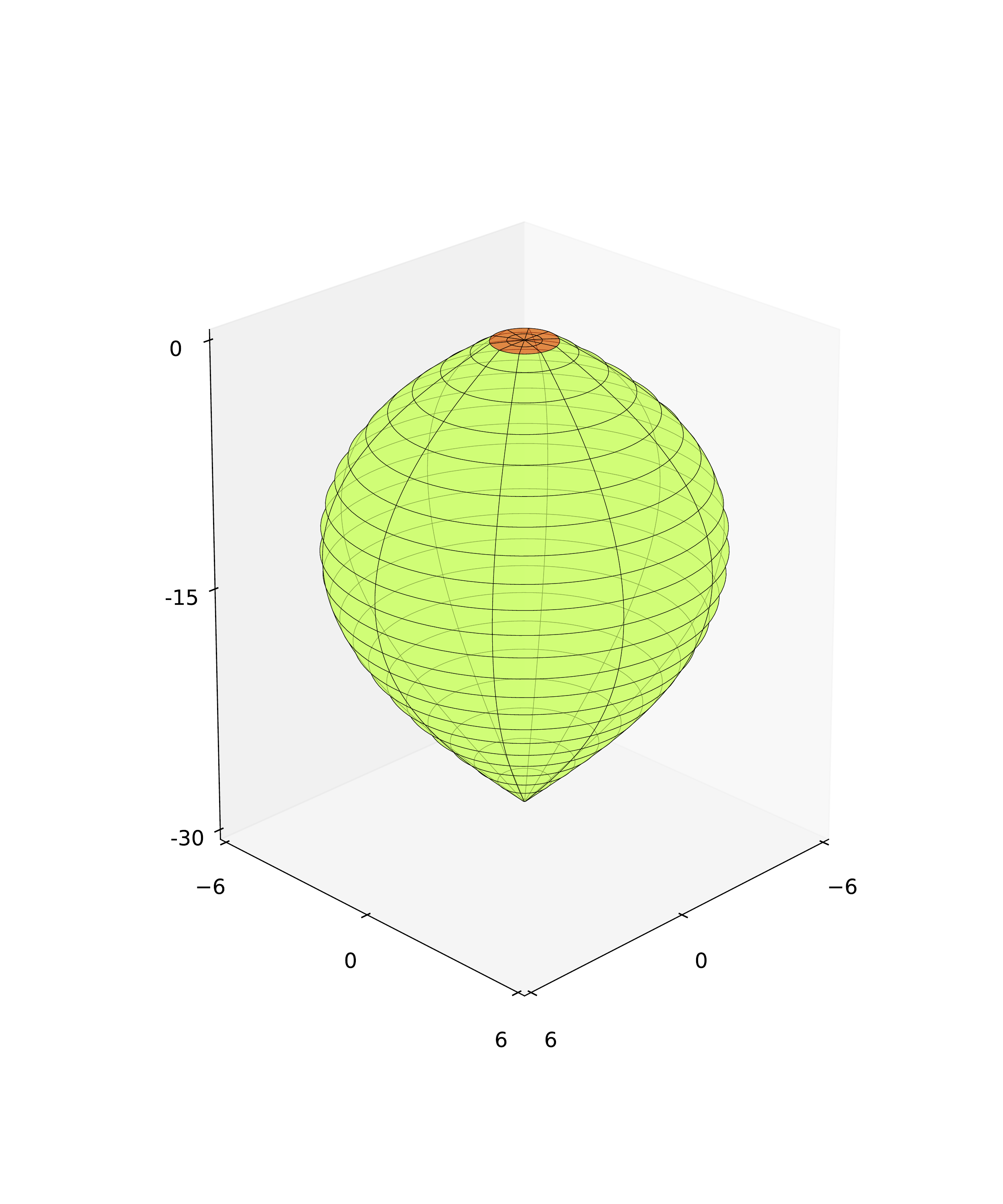}
	\caption{Embedding picture of the numerical solution obtained for the specific parameter choice~\eqref{eq:param} and $V= 256\, \pi $ (in units of the bulk Planck scale $ \kappa $). The regularized northern brane (which breaks SI) is localized along the ring separating the interior (red/dark) from the exterior (green/bright) region. The conical singularity at the south pole is caused by the unregularized (SI) pure tension brane.}
	\label{fig:rugby_ball}
\end{figure}

For simplicity, the brane at the south pole is chosen to be a pure tension brane without dilaton coupling, for which no regularization is required as it only leads to a conical defect of size
\begin{equation}
\alpha_-=1-\frac{\kappa^2}{2\pi}\mathcal{T}_- \,.
\end{equation}
The northern brane, which breaks SI, is regularized and now sits near the north pole at the coordinate position $ \rho_+ $, corresponding to a proper circumference $\ell \equiv 2 \pi B_+ >0$.\footnote{Here and henceforth, evaluation at $ \rho = \rho_0 $, $ \rho_+ $ and $ \rho_- $ will be denoted by subscripts ``$ 0 $'', ``$ + $'' and ``$ - $'', respectively.} The position of the (regular) axis at the north pole is denoted by $\rho_0$ $(<\rho_+)$. We can perform a shift of the $\rho$ coordinate such that $\rho_0=0$.
Figure~\ref{fig:rugby_ball} depicts the regularized bulk geometry for the exemplary parameter choice~\eqref{eq:param}. The interior of the ring (red/dark) is almost flat, whereas the exterior (green/bright) has the usual rugby ball shape.

Since the delta function $ \delta_+ \equiv \delta(\rho - \rho_+) $ is now localized at the position of the finite width ring, the regularized equations of motion are then formally identical to those presented in Sec.~\ref{sec:thin_brane}, apart from one crucial further modification: In order to prevent the ring from collapsing, it is necessary to introduce an angular pressure component, i.e.~to add the term
\begin{equation} \label{eq:add_p_theta}
	\frac{\delta_+}{2\pi B} \, p_\theta
\end{equation}
to the right hand side of the $ \comp{\theta}{\theta} $ Einstein equation~\eqref{eq:einstein_theta}. A possible way of modeling such a stabilization microscopically was first given in~\cite{Scherk:1978ta} and later also applied to the SLED model~\cite{Burgess:2008yx}: The idea is to introduce a localized scalar field that winds around the compact brane dimension and is subject to nontrivial matching conditions. As a result, shrinking the extra dimensions causes the related field energy to increase, hence implying a stable configuration with finite ring size. By integrating out the scalar field, it was explicitly shown in~\cite{Burgess:2008yx} that it contributes to the ($\phi$-dependent) tension on the brane and leads to a pressure in angular direction. The tension shift can be taken care of by an appropriate renormalization, and the whole stabilizing sector is then solely characterized by an angular pressure component $p_\theta$. 
Thus, without loss of generality, we will work with the renormalized theory. As argued in~\cite{Burgess:2008yx}, the value of $ p_\theta $ needed to stabilize the ring can be inferred from the Einstein equations.

The junction conditions across the brane can be readily derived and read\footnote{For convenience, here and throughout the rest of Sec.~\ref{sec:thick_brane}, we set $ W_+ = 1 $, which is always possible by a (rigid) rescaling of the 4D coordinates.}
\begin{subequations}\label{eq:matching}
	\begin{align}
	[B\phi']_{\rm disc}					&= \frac{\kappa^2}{2\pi}\mathcal{C}_+\,, \label{eq:jump_phip}\\
	4[B (\ln W)']_{\rm disc}	 				&=\frac{\kappa^2}{2\pi}\,p_{\theta} \,, \label{eq:jump_Wp}\\
	[B']_{\rm disc}					&=-\frac{\kappa^2}{2\pi}\left[ \mathcal{T}_+(\phi)+\frac{3}{4}\,p_{\theta}\right]_{\rho=\rho_+}\,, \label{eq:jump_Bp}
	\end{align}
\end{subequations}
where we introduced the notation
\begin{equation}
[f]_{\rm disc}:=\lim\limits_{\epsilon\to 0}\left[f(\rho_+ + \epsilon) - f(\rho_+ - \epsilon)\right]\,,
\end{equation}
for any function $ f(\rho) $. 

Furthermore, we have to impose appropriate boundary conditions at both axes. Since the north pole is regularized, the corresponding axis (at coordinate position $ \rho=0 $) is required to be elementary flat, i.e.
\begin{align}
\label{eq:bdry_N}
\phi'_0=0\,, && W'_0=0\,, && B'_0=1\,,&& B_0=0\,.
\end{align}
In general, the unregularized south pole  (at coordinate position $ \rho=\rho_- $) features a conical singularity characterized by
\begin{align}
\label{eq:bdry_S}
\phi'_-=0\,, && W'_-=0\,, && B'_-=-\alpha_-\,,&& B_-=0\,.
\end{align}
Note that only three of the four boundary conditions at each axis are independent, due to the radial Einstein constraint \eqref{eq:einstein_rho}. Let us now count the total number of integration constants: There are two second order and one first order equation, leading to a total of five a priori undetermined integration constants. In addition, there is one integration constant included in the metric ansatz \eqref{eq:ansatz_met}, namely $\hat R$.  All of them are fixed by imposing the six independent boundary conditions stated above. The closed system for $\phi$, $W$ and $B$ is thus given by the off-brane ($ \rho\neq\rho_b $) equations \eqref{eq:dilaton} and \eqref{eq:einstein_expl}, the junction conditions across the ring \eqref{eq:matching} and the boundary conditions \eqref{eq:bdry_N} and \eqref{eq:bdry_S} at the north and south pole, respectively. 

After fixing the above boundary conditions, we are left with a one-parameter family of solutions, parametrized by the Maxwell integration constant $ Q $. However, it cannot be chosen freely, because it contributes to the total flux $ \Phi_\mathrm{tot} := \int\!\rd\rho\,\rd\theta\, F_{\rho\theta} $, which is subject to the flux quantization condition~\cite{Randjbar:1983,Burgess:2011va},
\begin{equation} \label{eq:flux_quant}
	\Phi_\mathrm{tot} = 2\pi Q \int \!\rd \rho \, \frac{\re^{\phi} B}{W^4} + \left[ \mathcal{A}_+(\phi) \re^{\phi} \right]_{\rho=\rho_+} \stackrel{!}{=} \frac{2\pi n}{\tilde e} \qquad (n \in \mathds{N}) \,,
\end{equation} 
where  in general the U(1) gauge coupling $\tilde e$ can be different from $e$.

\subsection{4D Curvature}
The 4D curvature is crucial in studying the phenomenological viability of the model, so let us again derive its relation to the brane couplings, but now for the regularized model.
Repeating the derivation that lead to~\eqref{eq:degrav_cond_delta} in the thin brane setup, and taking into account~\eqref{eq:add_p_theta}, we now find
\begin{equation} \label{eq:R_hat_delta_2}
	V \hat R  = \kappa^2  \left( 2 \mathcal{C}_+ + p_\theta \right) \,.
\end{equation}
%
We see that the regularized expression is only modified by the last term proportional to $p_{\theta}$. Next, we will also express $ p_\theta $ in terms of the brane couplings in the thin brane limit.

\subsection{Angular Pressure and Delta Limit}
\label{sec:delta_limit}
The aim of this section is to explicitly check whether the above relations are compatible with the delta results of \cite{Niedermann:2015via}, and to gain further intuition about the regularized system and its stabilization. This will in turn allow us to narrow down physically interesting dilaton couplings.

Whether the brane looks pointlike to a good approximation is determined by the hierarchy between brane and bulk size, i.e.~by the dimensionless ratio $ \epsilon := \ell^2 / V $. Thus, the delta limit corresponds to $ \epsilon \to 0 $, and can be realized by letting $ \ell \to 0 $ and/or $ V \to \infty $.
In this work, we will keep $ \ell $ fixed at a value not smaller than the bulk Planck length,\footnote{Specifically, we will set $ \rho_+ = \sqrt{\kappa} $ in the numerical examples below, corresponding to $ \ell \approx 2 \pi \sqrt{\kappa} $.} and let $ V $ become large.

Let us first check whether the matching conditions \eqref{eq:matching} are compatible with the delta results~\eqref{eq:matching_delta} in the limit $\epsilon \to 0$. Since the geometry is close to flat space in the vicinity of the regularized axis, we assume\footnote{These assumptions were also verified numerically. }
\begin{align} \label{eq:rho_deriv_in}
\lim_{\rho\nearrow\rho_+}\phi' = \mathcal{O}(\epsilon) \,, && \lim_{\rho\nearrow\rho_+}W' = \mathcal{O}(\epsilon)\,, && \lim_{\rho\nearrow\rho_+}B' = 1 + \mathcal{O}(\epsilon) \,.
\end{align}
In that case, 
Eq.~\eqref{eq:jump_phip} indeed reduces to the dilaton boundary condition~\eqref{eq:jump_delta_phi} as $ \epsilon \to 0 $. On the other hand, Eqs.~\eqref{eq:jump_Wp} and~\eqref{eq:jump_Bp} show that the boundary conditions for $W$ and $B$ are again modified by a term proportional to $p_{\theta}$. This was also observed in \cite{Burgess:2008yx}.

At this point several remarks are in order:

\begin{itemize}
	
	\item The delta results  \cite{Niedermann:2015via} are recovered if and only if $\lim\limits_{\epsilon \to 0} p_\theta =0$.
	
	\item The occurrence of $p_\theta$ is expected, and a mere consequence of regularizing the setup as a ring. It has the clear physical interpretation as the angular pressure that is needed to stabilize the compact dimension.
	
	\item From a physical perspective, there is no understanding of an angular pressure for an infinitely thin object. As a result, we expect the pressure to vanish whenever there is a large hierarchy between the bulk size $V$ and the regularization scale $\ell$. 
	This expectation is in accordance with the above observation that for $p_\theta \to 0$ all results of the delta analysis are recovered. 
	Our present analysis allows to go beyond physical expectations and to explicitly take the thin brane limit.
	
	\item For the physically relevant class of exponential couplings (which admit a small 4D curvature and a large bulk volume), we will confirm the above expectation by showing $\lim\limits_{V \to \infty} p_\theta = 0$. This result also confirms the correctness of the delta approach in~\cite{Niedermann:2015via} within this class of couplings. While it is possible to construct examples in which $ p_\theta \nrightarrow 0 $, these are typically plagued by some sort of pathology, like a runaway behavior or a diverging brane energy (cf.~Sec.~\ref{sec:near_SI}). Again, this is not very surprising, as there is no  meaningful notion of a pointlike angular pressure.
	
	\item The authors of~\cite{Burgess:2015kda} instead argued that $ p_\theta $ should be nonzero for SI breaking delta branes. We comment on this in Appendix~\ref{ap:Cliff}.
	
\end{itemize}

We will now derive an expression for $p_\theta$ in terms of the dilaton coupling. This in turn enables us to identify and discuss those couplings that are compatible with the delta description. As we will see, these are also just the ones that lead to small $ \hat{R} $.

As pointed out in \cite{Burgess:2008yx}, an expression for $p_\theta$ can be found by evaluating the radial Einstein constraint \eqref{eq:einstein_rho} in the limit $\rho \searrow \rho_+$:
\begin{multline}	
3\left( \kappa^2 p_{\theta} \right)^2 - 8 \left(2 \pi - \kappa^2 \mathcal{T}_+ \right) \kappa^2 p_\theta + 4 \kappa^4\, \mathcal{C}^2_+\\
-\epsilon\, 8 V \hat R + \epsilon\, 4 \kappa^2 V \re^{\phi_+}  \left( Q^2 - \frac{4 e^2}{\kappa^4} \right) = \mathcal{O}(\epsilon)\,,
\end{multline}
where we used \eqref{eq:rho_deriv_in} and \eqref{eq:matching} to express the radial derivatives through the brane fields.
The terms in the second line are suppressed by $\epsilon$ and can be neglected in the delta limit. Solving for $ p_\theta $, we find
\begin{align}
\label{eq:ptheta1}
\kappa^2 p_\theta = \frac{4}{3}\left\{ \left(2\pi-\kappa^2 \mathcal{T}_+ \right)\pm \sqrt{ \left(2\pi-\kappa^2 \mathcal{T}_+ \right)^2-\frac{3}{4} \kappa^4\, \mathcal{C}^2_+}\right\} + \mathcal{O}(\epsilon)\,
\end{align}
where the branch was chosen such that the delta result $ p_\theta = 0 $ is recovered for SI couplings in the limit $ \epsilon \to 0 $.\footnote{Note that we only consider \emph{subcritical} tensions $ \mathcal{T}_+ < 2\pi / \kappa^2 $.} For vanishing BLF this coincides with the result derived in~\cite{Burgess:2008yx}.

An important observation from the above equation is that for finite $\epsilon$ and SI couplings in general\footnote{\label{fn:no_warp}There is a special class of SI solutions with $W'=0$ (no warping), $Q=2e/\kappa^2$ and $\hat R=0$ for which $p_\theta=0$ as an exact result even for $\epsilon\neq 0$. Physically, these solutions correspond to the regularized rugby ball setup. However, with respect to the CC problem this class is of no interest as it requires to unacceptably tune the relative size of both tensions.} $p_\theta = \mathcal{O}(\epsilon)\neq 0$. The physical reason is that introducing a brane width in general requires a stabilizing angular pressure.

The requirement of being close to SI can be made more precise by defining a near SI regime according to
\begin{align}
\label{eq:scale_invariant_limit}
\kappa^2 \mathcal{C}_+ \ll 1 \,.
\end{align}
This in turn leads to an approximate expression for the stabilizing pressure,
\begin{align}
\label{eq:ptheta2}
p_\theta = \frac{\kappa^2}{4\pi} \left(1-\frac{\kappa^2 \mathcal{T}_+}{2 \pi} \right)^{-1} \mathcal{C}^2_+ + \mathcal{O}(\epsilon) +  \mathcal{O}(\mathcal{C}_+^4) \,.
\end{align}
After inserting this into the formula for $\hat R$ in \eqref{eq:R_hat_delta_2}, we arrive at
\begin{equation}
\label{eq:R_hat_delta_3}
\boxed{
	V \hat R  = 2 \kappa^2\,\mathcal{C}_+ + \frac{1}{4\pi} \left(1-\frac{\kappa^2 \mathcal{T}_+}{2 \pi} \right)^{-1} \kappa^4\,\mathcal{C}^2_+  +\mathcal{O}(\epsilon) + \mathcal{O}(\mathcal{C}_+^4)
}\,.
\end{equation}
By comparing to its delta counterpart~\eqref{eq:degrav_cond_delta}, we find two small corrections:
\begin{enumerate}[(i)]
	\item a term quadratic in $\mathcal{C}_+$ and hence suppressed (in the near SI regime) relative to the leading linear term; 
	\item generic order $\epsilon$ contributions caused by the finite brane width.
\end{enumerate}
Which of the two dominates depends on the details of the dilaton coupling. Later, we will find that both possibilities can be realized.

In summary, we have shown that the delta result for $\hat R$ receives two corrections which are small in the near SI regime (which we intend to study) and for a large hierarchy between the brane size and extra space volume.

\subsection{Modeling Near Scale Invariance}
\label{sec:near_SI}
As expected, the near SI regime is of superior phenomenological importance as it leads to parametrically small values of the 4D curvature due to~\eqref{eq:R_hat_delta_3}.
We look for a dilaton coupling which allows to keep the SI breaking effects small without introducing an a priori hierarchy of the coupling parameters. In principle, this can be realized by using exponential couplings~\cite{Burgess:2015nka,Burgess:2015lda}, i.e.\
\begin{align} \label{eq:BLFcoupling}
	\mathcal{T}_+(\phi) = \lambda_+ + \tau\, \re^{\gamma\phi} &&\text{and} &&	\mathcal{A}_+(\phi) = \Phi_+ \re^{- \phi}\, ,
\end{align}
with $\phi$-independent (and SI) tension $\lambda$ and constant parameters $ \gamma $, $ \tau $ and $\Phi_+$. For $ \tau $ and $ \gamma \neq 0$ the tension term breaks SI explicitly. We see that even for (a naturally) large $\tau$, the SI breaking given by $\mathcal{T}'_+$ becomes small when $\phi_+$ is sufficiently negative. This makes the exponential couplings interesting with respect to the CC problem.

By contrast, the BLF term preserves SI. Technically, we could have introduced the SI breaking also via the BLF term, which would lead to the same outcome.\footnote{In fact, we checked this explicitly. The reason is that the terms $\mathcal{T}'_+$ and $(\re^{\phi}\mathcal{A}_+)'$  (which lead to SI breaking if nonvanishing) always occur in the combination \eqref{def:SI_combination}, so technically it makes no difference which of the two mediates the SI breaking.} However, it should be noted that it is physically more imperative to include a SI breaking tension as we expect loops of localized brane matter, which in general breaks SI,\footnote{A SI matter theory would lead to observational problems: As argued in~\cite{Burgess:2015lda}, this would imply a direct coupling between brane matter and $\phi $, corresponding to an additional (Brans-Dicke like) force of gravitational strength. This is clearly ruled out by solar system observations~\cite{Will:2005va} unless a mechanism is included to shield the dilaton fluctuations inside the solar system. A complete study of this case is thus beyond the scope of our present work.} to contribute to $\tau$. In other words, there is no obvious way of having $\tau$ small without imposing a fine-tuning. As a consequence, when looking for natural solutions, we have to consider a $\phi $-dependent tension with generic coefficient $\tau$. On the other hand, in the case of the BLF term, it depends on the details of the matter theory whether we expect loop corrections to $\Phi_+$. Following the discussion in \cite{Burgess:2015lda}, if the matter fields are not coupled directly to the Maxwell sector, there might be a chance of keeping SI breaking contributions to $\mathcal{A}_+$ small. In any case, including a breaking via the BLF term would, due to \eqref{eq:R_hat_delta_3}, yield an additional contribution to $\hat R$ and, as we will see, would make it even more difficult to comply with the observational constraints.

With these couplings we find
\begin{equation} \label{eq:C_near_SI}
\mathcal{C}_+ = \tau \gamma\, \re^{\gamma\,  \phi_+} \,,
\end{equation}
leading to an angular pressure
\begin{align}
\label{eq:ptheta3}
p_\theta =  \frac{\kappa^2}{4\pi \alpha_+} \left(\tau \gamma \re^{\gamma \phi_+}\right)^2+ \mathcal{O}(\epsilon) + \mathcal{O}(\mathcal{C}_+^3)\,,
\end{align}
where $ \alpha_+:= 1-\frac{\kappa^2}{2\pi}\lambda_+ $.
The numerical analysis we conduct in this work (cf.\ Sec.\ \ref{sec:num_results}) will show emphatically that the volume obeys\footnote{In the special case of a scale invariant coupling ($\gamma=0$) and delta branes, this follows analytically from the GGP solutions~\cite{Gibbons:2003di}, see~\cite{Niedermann:2015via}. } 
\begin{align} \label{eq:volume_scaling}
V \propto \re^{-\phi_+} \,,
\end{align}
hence implying 
\begin{equation} \label{eq:p_theta_scaling}
p_\theta \propto 
\begin{cases}
V^{-2\gamma} & \qquad (\text{for}\; 0 < \gamma <1/2) \\
V^{-1}  & \qquad (\text{for}\; \gamma=0 \text{ or } \gamma > 1/2)
\end{cases} ,
\end{equation}
asymptotically for $ V/\kappa \gg 1 $. The second line follows from the observation that for $\gamma>1/2$ the first expression in \eqref{eq:ptheta3} becomes sub-dominant compared to the $\mathcal{O}(\epsilon)$ contribution. The case $\gamma=0 $ is special as it corresponds to a SI coupling, where SI is only broken by the regularization. From \eqref{eq:ptheta3} it is clear that it is not continuously connected to $\gamma \neq 0 $ because the first term vanishes identically (irrespective of the value of $V$). In both cases, $\gamma=0 $ and $\gamma>1/2 $, the exponent saturates to the constant value $-1$.

The above formula allows us to discuss the consistency of the delta limit. We distinguish two cases:

\begin{enumerate}
	\item For $\gamma \geq 0$, increasing the volume of the compact space leads to a decreasing angular pressure. In other words, when we make the hierarchy between transverse brane size and bulk volume large, the angular pressure tends to zero in accordance with the physical expectation. Moreover, in this limit the SI case is approached (since $ \C_+ \propto \gamma V^{-\gamma} \to 0 $), which renders the above approximations more and more accurate. As an aside, note that this observation, i.e.\ the concurrency of $p_\theta$ being small and having a small amount of SI breaking, is the loophole to the objections raised in~\cite{Burgess:2015kda}. We discuss this more extensively in Appendix~\ref{ap:Cliff}.
	
	\item For $ \gamma < 0 $ the situation is different: If $\tau>0$, the system eventually hits a point (just before it becomes super-critical) where \eqref{eq:ptheta1} yields no real solution for $ p_\theta $ anymore, indicating a runaway behavior. Therefore, a discussion of that case requires the inclusion of a general time dependence of the fields which is beyond the scope of this work.

	On the other hand, if $\tau<0$, there are static solutions for which $p_\theta$ grows as $V$ is increased  due to \eqref{eq:ptheta1}.  This is related to the observation that the system gets driven away from SI ($ \C_+ \to \infty $). As a result, the 4D curvature $\hat R$ cannot be kept under control for a phenomenologically large $V$ unless the coefficient $\tau$ is tuned to be extremely small. Moreover, the tension tends to $- \infty $ in this case which strongly questions the physical consistency of these solutions. So this case is not interesting, neither phenomenologically nor with respect to the tuning issue. 
\end{enumerate}

In summary, the exponential coupling with $\gamma \geq 0$ is of particular interest, as it allows to be close to SI, which is important to make the 4D curvature parametrically small. This is achieved by considering a sufficiently large bulk volume. Other types of couplings (including monomial and exponential ones with $ \gamma <0$) either lead to a runaway behavior or are incompatible with being close to SI (if the coefficient is not tuned to be small). The above discussion also shows that the physically relevant class of couplings is compatible with the delta description because $p_\theta$ (or any hidden metric dependence of the delta function as argued in \cite{Burgess:2015kda}) vanishes for $V \to \infty$.

\subsection{Phenomenology} \label{sec:phen}

We have singled out the exponential tension-dilaton coupling \eqref{eq:BLFcoupling} as the phenomenologically relevant one, since its contribution to the 4D curvature can be made arbitrarily small. Let us now discuss whether this can lead to phenomenologically viable solutions.

At the present stage, there are two main phenomenological inputs the model has to comply with:

\begin{enumerate}[(1)]
	\item In models with large extra dimensions the weakness of 4D gravity is a result of the large extra dimensions. This is possible because the 4D Planck mass is given, via dimensional reduction, by~\cite{Burgess:2015lda}
	\begin{align} \label{eq:Mp_phen}
	\Mp^2 = \frac{V}{\kappa^2}\,.
	\end{align}
	%
	
	Given present tests of the gravitational inverse square-law~\cite{Kapner:2006si} (see~\cite{Adelberger:2003zx} for a review), the upper bound on the size of the extra dimensions is of order of ten microns. Then, \eqref{eq:Mp_phen} implies that the bulk gravity scale  $\kappa^{-1/2}$ is not allowed to be significantly below $ \sim 10 \;\mathrm{TeV} $, which translates into the upper bound
	\begin{align}\label{eq:V_phen}
	\frac{V}{\kappa} \lesssim 10^{28} \,.
	\end{align}

	\item The observed value of the 4D curvature measured in Planck units is notoriously small, viz.~\cite{Ade:2013zuv}
	\begin{align} \label{eq:R_phen}
	\frac{\hat R}{\Mp^2} \sim 10^{-120}\,.
	\end{align}  
\end{enumerate}

Let us now study whether the model is compatible with both requirements. For convenience, we will set  $ \kappa = 1 $, i.e.\ here and henceforth dimensionful quantities are all measured in units of the bulk gravity scale.

We now make use of our central formula \eqref{eq:R_hat_delta_3} which permits to express the 4D curvature in terms of the extra space volume. Using \eqref{eq:C_near_SI}, \eqref{eq:volume_scaling} as well as \eqref{eq:Mp_phen}, we then find
that the leading contribution is
\begin{equation} \label{eq:R_scaling}
	\frac{\hat R}{\Mp^2} = N_1 V^{-(2+\gamma)} + N_2 V^{-3} \,,
\end{equation}
where $ N_i $ are dimensionless coefficients, with
\begin{align} \label{eq:N12}
	N_{1} \propto \gamma\tau && \text{and} && N_2 \propto \ell^2 \,.
\end{align}
The unknown constants of proportionality are due to the unknown coefficients in~\eqref{eq:volume_scaling} and the $ \mathcal{O}(\epsilon) $ term in~\eqref{eq:R_hat_delta_3}, respectively. 
For model parameters which do not contain a priori hierarchies among themselves, we expect them to be roughly $ \sim 1 $. While at this point it is merely a reasonable expectation, it will also be confirmed by the numerical solutions discussed in Sec.~\ref{sec:num_results}, which allow us to explicitly calculate these coefficients.
The relation~\eqref{eq:R_scaling} is one of the main results of this work.
The two phenomenological bounds above then require
\begin{equation}
	N_1 \times 10^{-28(2+\gamma)} + N_2 \times 10^{-84} \lesssim 10^{-120}\,.
\end{equation}
One way how this could in principle be fulfilled is by assuming a cancellation of the two terms. However, this would only be achieved by tuning the parameters $ \gamma $ and $ \tau $ very accurately. Therefore, we dismiss this possibility and demand both terms to fulfill the bound separately.
From~\eqref{eq:N12} we know that the first term vanishes identically for a SI coupling ($ \gamma\tau = 0 $). If SI is broken, it could only comply with the bound without tuning $ N_1 $ (and thus $ \tau $) if $ \gamma \gtrsim 2.3 $.\footnote{In Sec.~\ref{sec:num_results}, however, we will uncover yet another fine-tuning (imposed by flux quantization) which could only be avoided if $ \gamma \ll 1 $.} The second term, however, is more problematic: it implies that $ N_2 \lesssim 10^{-36} $. As expected from~\eqref{eq:N12}, and explicitly confirmed in Sec.~\ref{sec:num_results}, this could only be achieved by assuming the brane width $ \ell $ to be $ \sim 18 $ orders of magnitude smaller than the bulk Planck length. Not only would this again correspond to introducing an a priori hierarchy by hand, but also question the applicability of a classical analysis.

As a result, \emph{if we do not allow the model parameters to be fine-tuned or to introduce large hierarchies, the model is ruled out phenomenologically}. Either the 4D curvature or the size of the extra dimensions would be too large to be phenomenologically viable.


Before concluding this sections, let us summarize the assumptions that went into this result:

\begin{itemize}
	\item The interior profiles are close to their flat space estimates with corrections $\mathcal{O}(\epsilon)$, cf.~Eq.~\eqref{eq:rho_deriv_in}.
	
	\item Motivated by the GGP result, the extra space volume is assumed to be proportional to $ \re ^{-\phi_+}$, cf.~Eq.~\eqref{eq:volume_scaling}.
	
	\item The coefficients in~\eqref{eq:N12} are of order unity.
\end{itemize}

They are all quite reasonable, and will indeed all be explicitly confirmed by our numerical analysis. Moreover, the numerical treatment will allow us to infer the amount of tuning (due to flux quantization) that is required to get a sufficiently small 4D curvature (albeit corresponding to a too large $ V $).

\section{Numerical Results and Fine-Tuning} \label{sec:num_results}

In this section we present the results of our numerical studies of the regularized model and discuss their physical implications for the SLED scenario. We will first briefly sketch the numerical algorithm in Sec.~\ref{sec:num_alg}. Next, in Sec.~\ref{sec:SI}, we will discuss the simple case of SI brane couplings. In this case we know the exact analytic solutions for infinitely thin branes---the GGP solution, reviewed in~\cite{Niedermann:2015via}---and so this provides a useful consistency check for our numerical solver. Finally, Sec.~\ref{sec:non_SI} addresses the actual case of interest: a SI breaking tension. We derive the solutions of the full brane-bulk system without relying on any approximations, which in turn enables us to explicitly test (and confirm) the analytical approximations and results of the last section.

\subsection{Numerical Algorithm and Parameters} \label{sec:num_alg}

The goal is to determine the $ \rho $-profiles of the dilaton $ \phi $ and of the metric functions $ B $ and $ W $ for given model parameters.
As explained above, this requires solving the bulk equations~\eqref{eq:dilaton}, \eqref{eq:einstein_expl}, supplemented by the junction conditions~\eqref{eq:matching} and the boundary conditions~\eqref{eq:bdry_N}, \eqref{eq:bdry_S}. We do so by starting at the north pole ($ \rho = 0 $) and integrating outward using the second order equations.\footnote{We used two independent implementations: one in Python, using an explicit adaptive Runge-Kutta method, and one in Mathematica, using its ``NDSolve'' method. The corresponding results were found to agree within the numerical uncertainties.} Since the constraint~\eqref{eq:einstein_rho} is analytically conserved, it only needs to be imposed initially at $ \rho=0 $. For $ \rho > 0 $ it can then be used as a consistency check (or error estimator) of the numerical solution. At $ \rho=\rho_+ $, however, the constraint must be used once again, because it determines the stabilizing pressure $ p_\theta $. In other words, when the integration reaches $ \rho \nearrow \rho_+ $, the three junction conditions~\eqref{eq:matching} must be supplemented by the constraint (evaluated at $ \rho \searrow \rho_+ $) in order to determine the three exterior $ \rho $-derivatives and $ p_\theta $.
Afterwards, the integration continues until $ B \to 0 $, defining the south pole $ \rho = \rho_- $. 

Before the equations can actually be integrated in this way, we need to specify the three a priori unknown integration constants $ \phi_0 $, $ Q $ and $ \hat{R} $. In general, however, all of them are ultimately fixed via (the SI case is exceptional, see Sec.~\ref{sec:SI})
\begin{enumerate}[(i)]
	\item flux quantization~\eqref{eq:flux_quant}, 
	\item regularity at the south pole, i.e.,~$ \phi'_- = 0 $,\footnote{The corresponding regularity condition for $ W $ is not independent thanks to the constraint, i.e.,~$ W'_- = 0 $ automatically whenever $ \phi'_- = 0 $.} 
	\item the correct conical defect at the south pole, i.e.,~$ B'_- = -\alpha_- $. 
\end{enumerate}
Technically, this can be achieved by a standard shooting method: we choose some initial guesses for $ \phi_0 $, $ Q $ and $ \hat{R} $; after integrating the ODEs, the violations of (i)--(iii) can be computed, and finally be brought close to zero via an iterative root-finding algorithm.

In this way---and in agreement with the discussion in Sec.~\ref{sec:ring_reg}---since there are no integration constants left (in the non SI case), we also see that the full solution is uniquely determined for a given set of model parameters. These consist of the bulk couplings $ \kappa = 1$ (in our present units), $ e $, the regularization width $ \rho_+ $, the brane couplings, parametrized by $ \alpha_\pm $, $\tau$, $\gamma$ and the BLF parameter $ \Phi_+ $, as well as the gauge coupling $ \tilde e $. Since the latter only enters via flux quantization~\eqref{eq:flux_quant}, it is convenient to introduce the abbreviation
\begin{equation}
\N:= \frac{2\pi n}{\tilde e} \,,
\end{equation}%
so that flux quantization simply reads $ \Phi_\mathrm{tot} = \N $.

Note that the solution would \emph{not} be determined uniquely if, for instance, the boundary conditions ensuring regularity at the south pole were neglected. In this case, it would not be possible to numerically predict the value of $ \hat{R} $, since it could be chosen freely. Thus, in order to \emph{compute} this quantity numerically, it is crucial to find complete, regular bulk solutions. To our knowledge, this is done here for the first time.\footnote{Analytically, the regularity condition also implicitly entered the derivation of~\eqref{eq:R_hat_delta_3} when integrating over the whole bulk. However, this equation for $ \hat{R} $ is not yet a prediction solely in terms of model parameters, since it still contains $ V $ and $ \phi_+ $, which are a priori unknown. We were only able to infer the explicit value of $ \hat{R} $ numerically.}

The main question is whether it is possible to find solutions for which $ \hat R $ is small enough and $ V $ is large enough to be phenomenological viable without fine-tuning, i.e.~for generic values of the model parameters. For definiteness, and in order not to introduce any large hierarchies into the model by hand, we will choose the following parameters,
\begin{align} \label{eq:param}
e = 1\,, && \rho_+ = 1\,, && \Phi_+ = -0.6\,, && \tau = 0.9\times 2 \pi\,, &&\alpha_+= 0.9 &&  \text{and} && \alpha_- = 0.5 \,.
\end{align}
(Somewhat different values would not change the main results, though.) The parameter $ \N $, determining the total flux, will be varied, and used as a dial to achieve different values of $ \hat{R} $ and $ V $. 

\begin{figure}[t]
	\includegraphics[width=\textwidth]{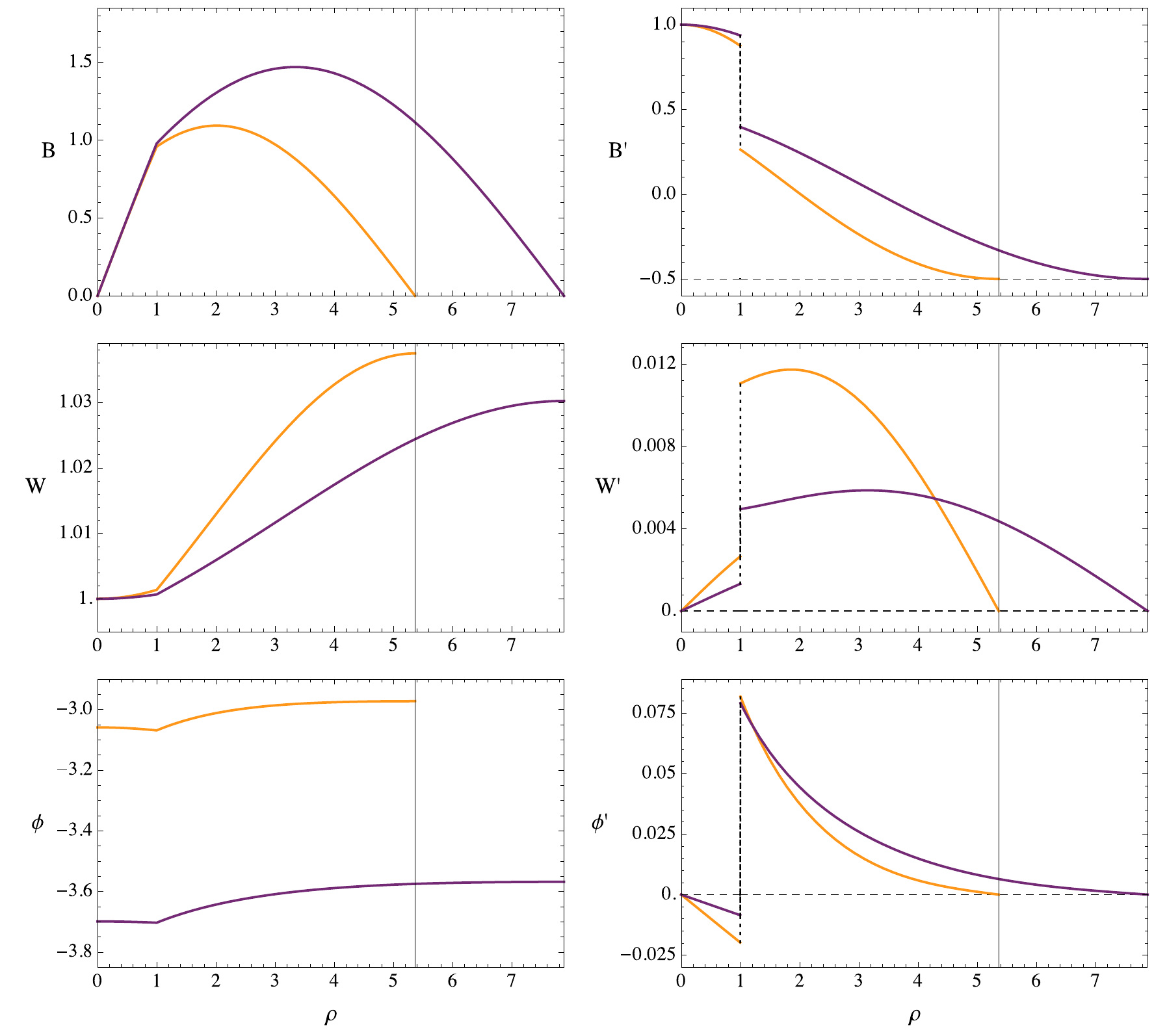}
	\caption{Complete numerical solutions of the coupled Einstein-dilaton system for the parameters~\eqref{eq:param} and $ \gamma = 0.2 $. The axis at the north pole ($ \rho = 0 $) is regular ($ W'=\phi'=0 $) and elementary flat ($ B'=1 $), while the axis at the south pole is regular but has a defect angle corresponding to the unregularized pure tension brane ($ B'=-0.5 $); the regularized brane sits at $ \rho_+ = 1 $, and produces jumps in the $ \rho $-derivatives. The orange (light) and purple (dark) curves correspond to $ V=8\pi $ and $ V = 16\pi $, respectively (which were obtained for $ \N = -1.102 $ and $ \N = -0.885 $). The required 4D curvature was $ \hat{R} = 0.0571 $ and $ 0.0233 $, respectively.
	The constraint violation, i.e.~the numerical deviation of~\eqref{eq:einstein_rho} from zero, was always smaller than $ 10^{-10} $ in this example, and the numerical error bars would not exceed the line widths in the plots. }
	\label{fig:profiles}
\end{figure}

An exemplary numerical solution is shown in Fig.~\ref{fig:profiles}, where the three functions $ B, W, \phi $, as well as their $ \rho $-derivatives are plotted, for $ \gamma = 0.2 $ and two different choices of $ \N $, leading to two different values of $ V $, as is evident from the profile of $ B $. Since we chose $ \alpha_+ \neq \alpha_- $, the solutions are warped---both $ W $ and $ \phi $ have nontrivial profiles.\footnote{Note that here we chose the gauge $ W_0 = 1 $ for convenience.}

Furthermore, one can already see that the profiles inside the regularized brane ($ \rho < \rho_+ $) become more trivial as $ V $ increases, as expected. This trend continues, and all functions and their derivatives at $ \rho \nearrow \rho_+ $ were always found to approach the corresponding values at the regular axis ($ \rho = 0 $) like $ V^{-1} $ for $ V \to \infty $, thereby confirming~\eqref{eq:rho_deriv_in}.

All of the $ \rho $-derivatives are discontinuous at the regularized brane ($ \rho = \rho_+ $), as required by the junction conditions~\eqref{eq:matching}. $ B' $ consistently approaches $ -\alpha_- = -0.5 $ at the south pole and, most importantly, both $ W' $ and $ \phi' $ vanish there, as required by regularity. By running the numerics similarly for different choices of $ \gamma $ and $ \N $, we can now systematically learn how these model parameters determine $ \hat{R} $ and $ V $.

\subsection{Scale Invariant Couplings and Thick Branes} \label{sec:SI}

Let us first consider the case $ \tau = 0 $ corresponding to a SI tension $ \mathcal{T}_+ = 2\pi(1-\alpha_+) $. Incidentally, in this case the dilaton profile is regular, and so the solution can even be obtained for the idealized, infinitely thin brane, as already discussed in~\cite{Niedermann:2015via}. It is given by the GGP solution~\cite{Gibbons:2003di}, for which $ \hat R = 0 $. In that case, the integral in the flux quantization condition~\eqref{eq:flux_quant} can be performed explicitly, yielding
\begin{equation} \label{eq:flux_quant_GGP}
	\frac{2 \pi}{e}\sqrt{\alpha_+\alpha_-} + \Phi_+ = \N \,.
\end{equation}
The dilaton integration constant $ \phi_0 $ drops out of all equations due to SI, and thus the above counting of constants does not add up, resulting in the tuning relation~\eqref{eq:flux_quant_GGP} among model parameters. If we chose parameters which do not fulfill this equation, there would not be a static solution, in accordance with the expected runaway behavior \`{a} la Weinberg~\cite{Weinberg:1988cp}.
In turn, the extra space volume $ V $, which turns out to be $ \propto \re^{-\phi_0} $~\cite{Niedermann:2015via}, can be chosen freely. As a result, this model could have a phenomenologically viable volume (although a vanishing 4D curvature is not compatible with observations), but only at the price of a new fine-tuning.

If SI is broken, things will change: on the one hand, $ \phi_0 $ will be fixed, and thus the tuning relation is expected to disappear. On the other hand, the volume $ V $ will also be determined, and $ \hat R $ is expected to be nonzero. The question then is if they can satisfy the phenomenological bounds presented in Sec.~\ref{sec:phen}, and if so, whether this can be achieved without introducing yet another tuning.

\begin{figure}[t]
	\subfloat[]{
		\includegraphics[width=0.475\textwidth]{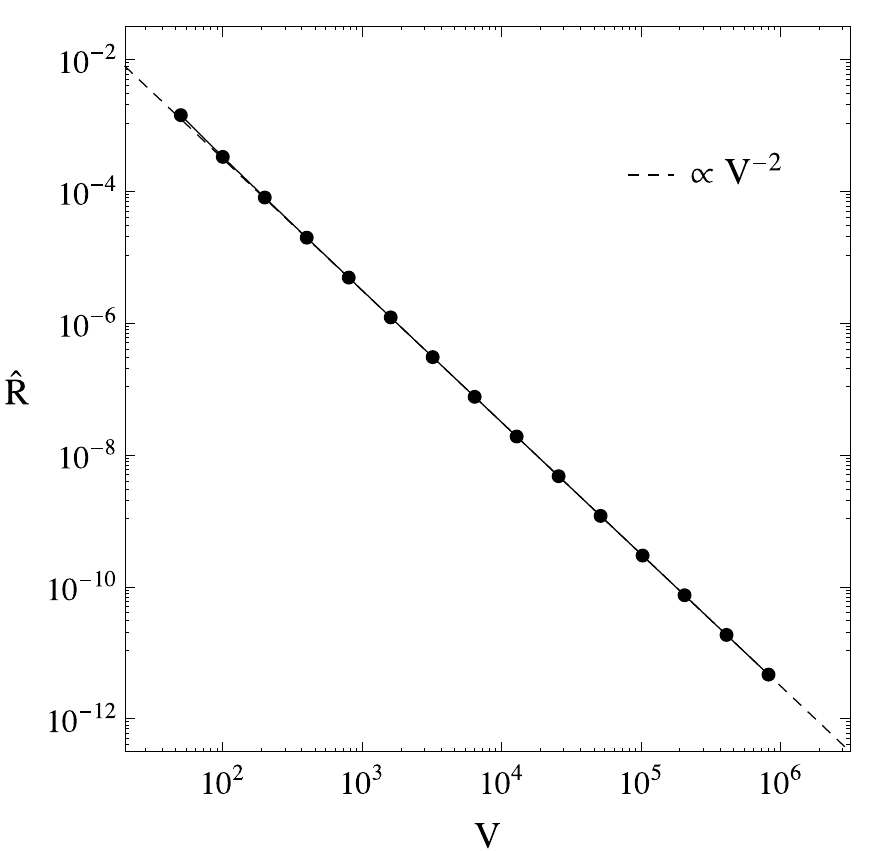}
		\label{fig:R0}
	}
	\hfill
	\subfloat[]{
		\includegraphics[width=0.475\textwidth]{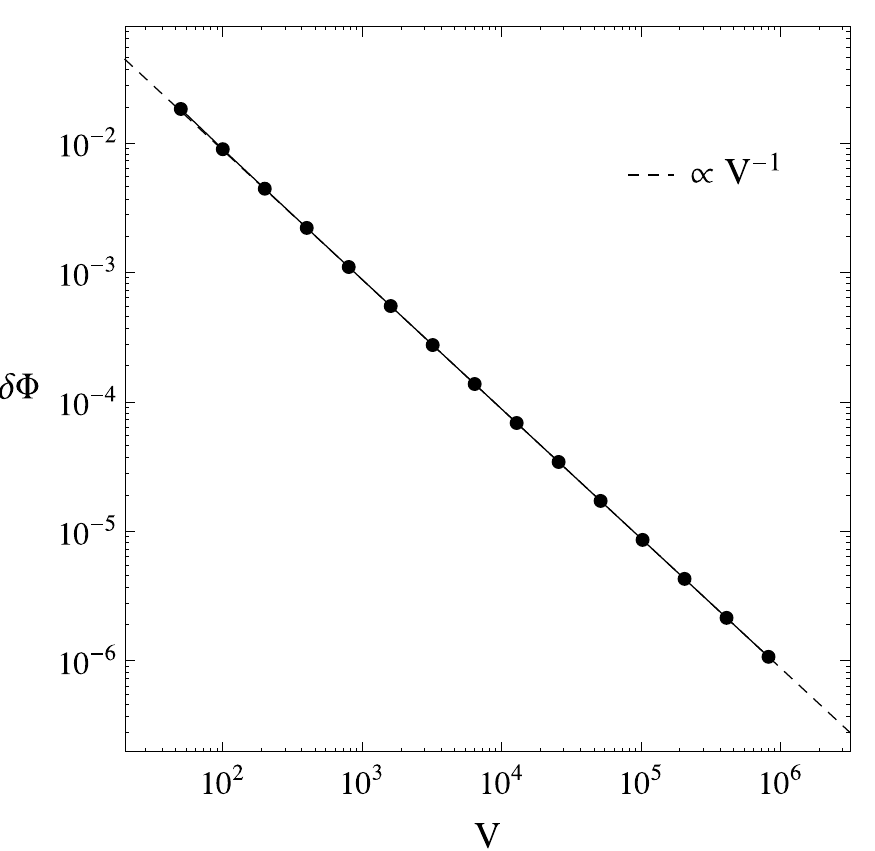}
		\label{fig:Flux0}
	}
	\caption{Numerical results for parameters~\eqref{eq:param} and $ \tau = 0 $, corresponding to SI brane couplings. For large volume $ V $, the 4D curvature and the total flux both approach the corresponding GGP values which are valid for delta branes. The dashed lines are numerically inferred (and extrapolated) scaling laws.}
	\label{fig:SI_coupling}
\end{figure}

Let us now present the numerical results for a regularized brane with $ \tau = 0 $ [all other parameters as in \eqref{eq:param}]. In that case SI is already broken by introducing a regularization scale $\ell$. Thus, the above discussion applies here as well: $ \phi_0 $ and $ V $ are fixed in terms of model parameters. Moreover, we expect $\hat R \neq 0$ due to $ \mathcal{O}(\epsilon)$ contributions caused by the finite brane width.\footnote{This is a qualitative difference to models with two \emph{infinite} extra dimensions, where a regularized pure tension brane still has $ \hat{R} = 0 $~\cite{Kaloper:2007ap,Eglseer:2015xla}.} However, if the thin brane limit is taken by letting $ V \to \infty $ (which can be achieved by adjusting $ \N $ appropriately), these effects should become suppressed, and we expect to recover the GGP solution with $ \hat{R} = 0 $. This is exactly what happens, as can be seen from Fig.~\ref{fig:R0}. Specifically, we find that $ \hat{R} \propto V^{-2} $ as $ V \to \infty $. Furthermore, the angular pressure $ p_\theta $ (not shown) is also nonvanishing, but goes to zero like $ V^{-1} $. These findings are in complete agreement with the analytic predictions~\eqref{eq:R_hat_delta_2}, \eqref{eq:ptheta1} (with $ \C_+ = 0 $).

At the same time, the tuning relation~\eqref{eq:flux_quant_GGP} is also violated, and the static solutions exist for any choice of parameters. But again this violation,
\begin{align}
\delta\Phi := \Phi_\mathrm{GGP} - \N \,, && \text{with} && \Phi_\mathrm{GGP} := \frac{2 \pi}{e}\sqrt{\alpha_+\alpha_-} + \Phi_+ \,,
\end{align}
vanishes (like $ V^{-1} $) as $ V \to \infty $, see Fig.~\ref{fig:Flux0}.\footnote{Incidentally, it turns out that without warping, i.e.~for $ \alpha_+ = \alpha_- $, the scalings are somewhat different: $ \hat{R} \propto V^{-3} $, $ \delta\Phi \propto V^{-2} $ and $ p_\theta \propto V^{-2} $. However, this does not help with the tuning problem discussed below.}

In summary, we explicitly confirmed that introducing a regularization leads to $ \mathcal{O}(\epsilon) $ corrections of the GGP predictions ($ \hat R = 0 $, $ \Phi_\mathrm{GGP} = \N $, $ p_\theta = 0 $). In particular, this agrees with the analytic result of~\cite{Niedermann:2015via} that $ \hat{R} = 0 $ is only guaranteed in the SI delta model (which is approached as $ \epsilon \to 0 $) via a tuning of model parameters ($ \Phi_\mathrm{GGP} = \N $). Furthermore, this simple example already shows that a stabilizing pressure $ p_\theta $ is necessary for a thick brane, but also that $ p_\theta \to 0 $ as $ \epsilon \to 0 $, allowing for a consistent delta description as in~\cite{Niedermann:2015via}.

But now we can even make a precise statement about the required tuning beyond the idealized delta brane limit. The phenomenological bound~\eqref{eq:R_phen} together with~\eqref{eq:Mp_phen} yields (recall that we are working in units in which $ \kappa = 1 $)
\begin{equation} \label{eq:RV_phen}
10^{-120} \stackrel{!}{\sim} \frac{\hat R}{V} \sim \delta\Phi^3 \,,
\end{equation}
where the second estimate used (and extrapolated) our numerically inferred scaling relations (neglecting the $ \mathcal{O}(1) $ coefficients), cf.~Fig.~\ref{fig:SI_coupling}. Therefore, the parameter $ \N \equiv 2\pi n / \tilde e $ must be tuned close to $ \Phi_\mathrm{GGP} \equiv \frac{2 \pi}{e}\sqrt{\alpha_+\alpha_-} + \Phi_+ $ with a precision of $ \sim 10^{-40} $. This is clearly not better than the CC problem we started with. It is crucial to note that this can also directly be read as a tuning relation for the brane tension $ \lambda $, since $ \alpha_+ = 1 - \lambda / 2\pi $.

But---as already anticipated in Sec.~\ref{sec:phen}---there is also another problem regarding phenomenology, even if we allow for such a tuning: For $ \delta\Phi \sim 10^{-40} $, the extra space volume would be $ V \sim 10^{40} $, grossly violating the bound \eqref{eq:V_phen}. Thus, by tuning $ \hat R $ small enough, we have at the same time tuned the extra space volume 12 orders of magnitude larger than allowed. Alternatively, if we require $ V $ to satisfy the observational bound~\eqref{eq:V_phen}, $ \hat{R} $ would still be 36 orders of magnitude larger than what is observed. Hence, as it stands, the model suffers not only from a tuning problem, but is not even phenomenologically viable.

This nicely agrees with the analytic discussion in Sec.~\ref{sec:phen}. Explicitly, we confirmed the relation~\eqref{eq:R_scaling} (here for $ \gamma = 0 $), finding the coefficient $ N_2 = 3.16 $ for this specific set of parameters, i.e.~$ e $, $ \rho_+ $, $ \Phi_+ $ and $ \alpha_\pm $ as given in~\eqref{eq:param}. Now, since the resulting failure to get both $ \hat{R} $ and $ V $ within their phenomenological bounds is the central result of this work, it is worthwhile to discuss its robustness.

First, it should be noted that the main reason for this result can be traced back to the $ \mathcal{O}(\epsilon) $ contributions to the 4D curvature $ \hat{R} $, cf.~Eq.~\eqref{eq:R_hat_delta_3}, which are caused by endowing the brane with a finite width. Hence, they are unavoidable in a (realistic) thick brane setup; of course, we did our explicit calculations only in one particular regularization, but the standard EFT reasoning suggests that the qualitative answer would be the same for any other reasonable regularization.\footnote{One could test this assumption by repeating our analysis e.g.~in the UV model proposed in~\cite{Burgess:2015nka}.} While there are additional contributions to $ \hat{R} $ if the dilaton couplings break SI, see Eq.~\eqref{eq:R_scaling}, they can only make things worse (unless there were a miraculous cancellation---a possibility that we dismiss in the search of a natural solution to the CC problem). Again, this will be explicitly confirmed in the following section.

Next, we checked numerically that the scaling relation, as well as the order of magnitude of the coefficient $ N_2 $ do not change if different tensions (i.e.~other generic values for $ \alpha_\pm $) are chosen. Furthermore, the parameters $ \Phi_+ $ and $ e $ have no influence on the result at all; this is obvious for the BLF $ \Phi_+ $, but also easily seen for the gauge coupling $ e $ as follows: For the SI couplings we are considering here, the full (regularized) equations of motion enjoy the exact symmetry
\begin{align} \label{eq:SI_symm}
	e \mapsto a e \,, && Q \mapsto a Q \,, && \re^{\phi} \mapsto \frac{1}{a^2} \re^{\phi} \,,
\end{align}
for any constant $ a $. Hence, after changing $ e $, the new solution is simply obtained from the old one by rescaling $ Q $ and $ \re^\phi $ appropriately. Since the metric is unaltered, this leaves $ \hat{R} $ and $ V $ unchanged.\footnote{Note that the (bulk) flux transforms as $ \Phi \mapsto \Phi / a $, and so $ \N $ has to be readjusted accordingly. This, however, does not affect the relation between $ \hat{R} $ and $ V $.}

Hence, the only parameter that could change things is $ \rho_+ $, determining the regularization scale $ \ell \approx 2\pi \rho_+ $, in accordance with the discussion below Eq.~\eqref{eq:N12}.

\subsection{Non Scale Invariant Couplings} \label{sec:non_SI}

We now turn to the case $ \tau \neq 0 $ (and $ \gamma > 0 $),\footnote{The case $ \gamma = 0 $ is still SI and identical to the discussion above after renaming $ \lambda + \tau \to \lambda $.} where SI is broken explicitly via the tension term. The hope is to find values of $ \gamma $ for which no tuning is required in order to achieve a large volume and small curvature. As argued above, this suggests focusing on $ \gamma > 0 $, because then $ V \to \infty $ drives the model towards the SI case which in turn implies $ \hat R \to 0 $. While this case was already discussed in Sec.~\ref{sec:phen} under certain reasonable assumptions, the numerical analysis independently confirms the previous results and allows to quantify the amount of tuning necessary to get a viable 4D curvature.

\begin{figure}[t]
	\subfloat[A small 4D curvature $\hat R$ is realized for a small violation $\delta \Phi$ of the GGP tuning relation, thus implying a highly tuned brane tension.]{
		\includegraphics[width=0.475\textwidth]{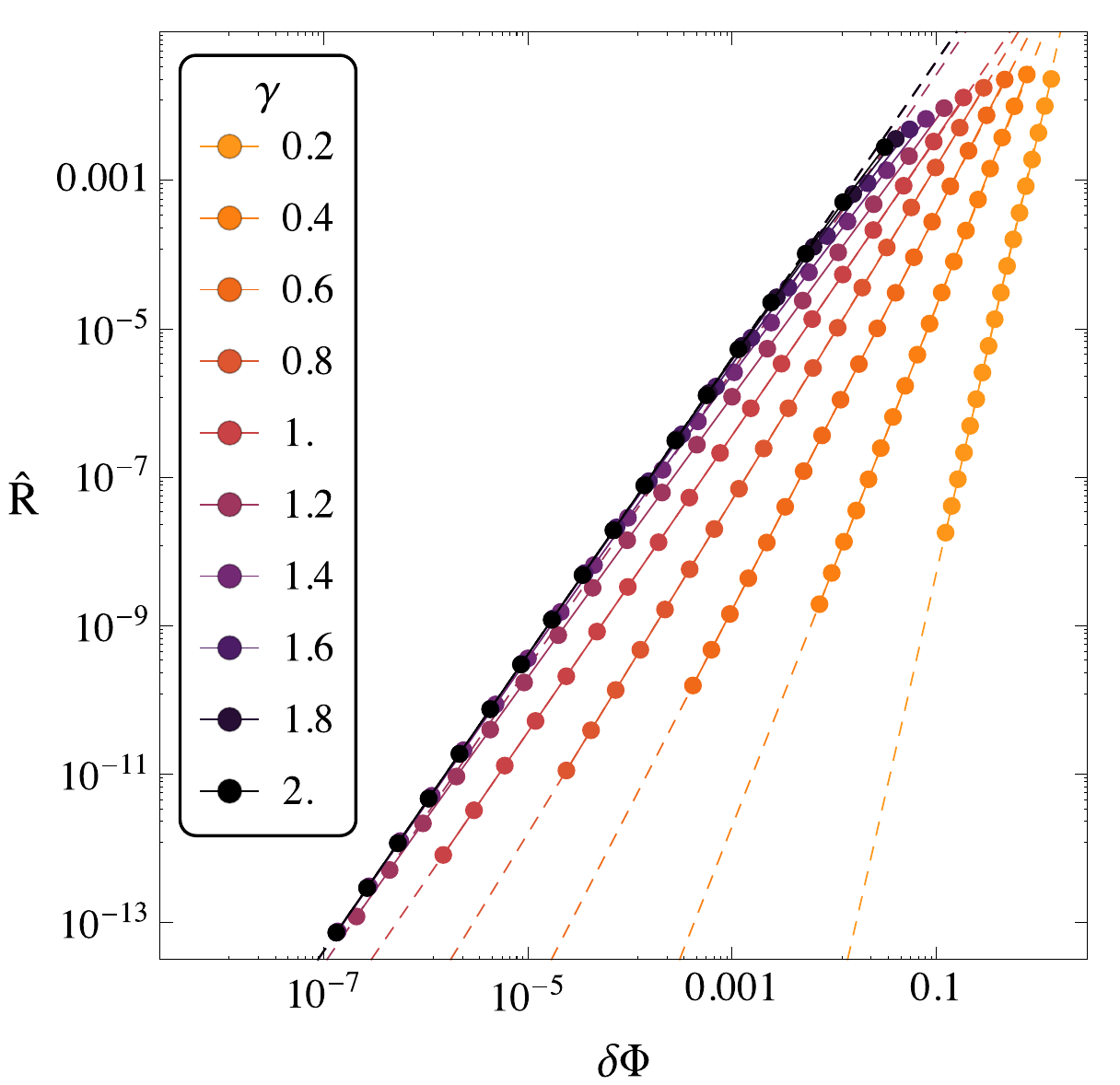}
		\label{fig:R}
	}
	\hfill
	\subfloat[A large extra space volume $ V $ is achieved for a small $\delta \Phi$.]{
		\includegraphics[width=0.475\textwidth]{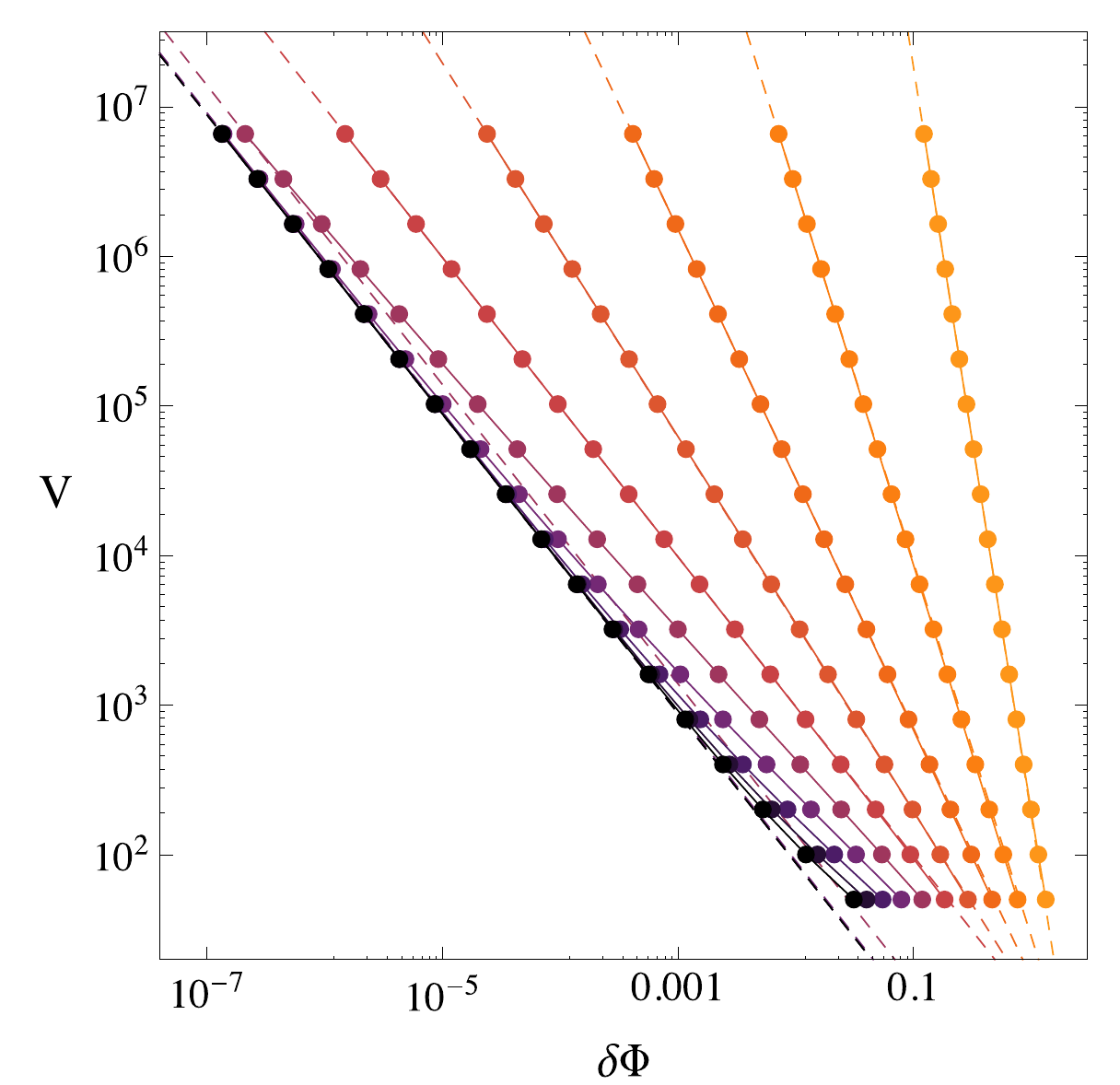}
		\label{fig:V}
	}\\
	\subfloat[The angular pressure $p_\theta$ vanishes in the thin brane limit in accordance with the EFT expectation.]{
		\includegraphics[width=0.475\textwidth]{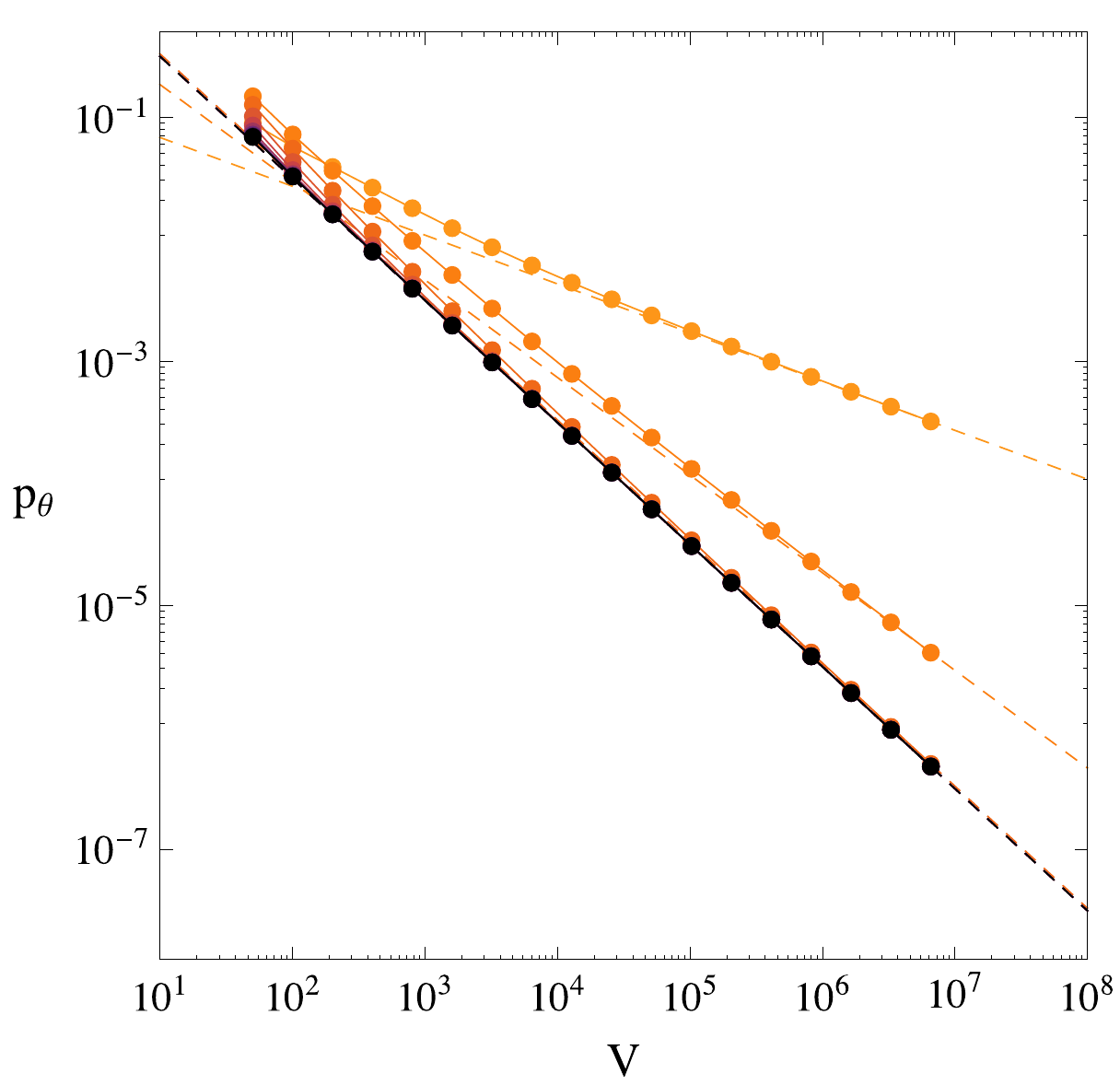}
		\label{fig:pTheta}
	}
	\hfill
	\subfloat[The dilaton evaluated at the brane $\phi_+$ controls the extra space volume $V$ via \eqref{eq:volume_scaling}.]{
		\includegraphics[width=0.475\textwidth]{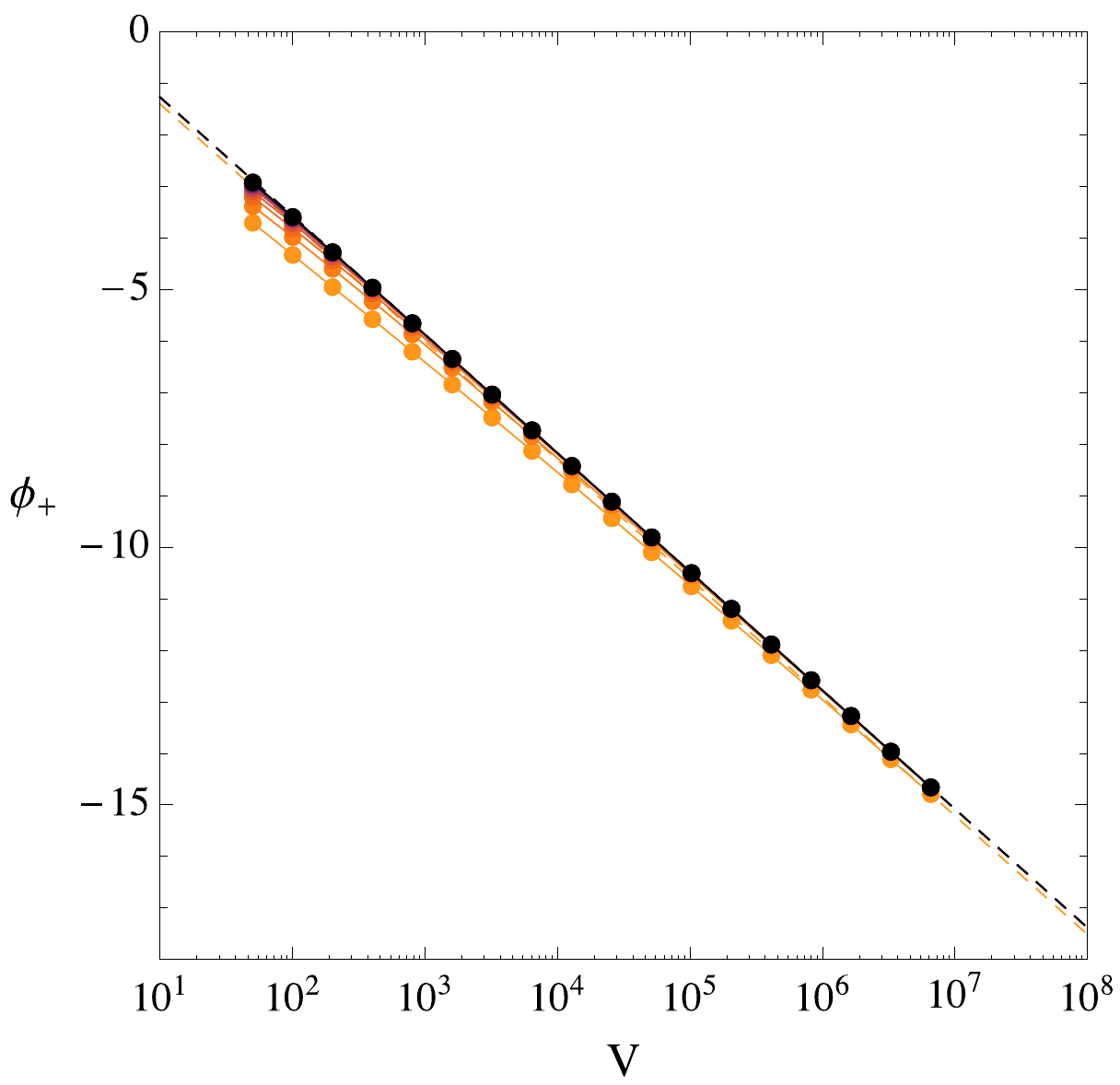}
		\label{fig:phi}
	}
	\caption{Numerical results for the parameters~\eqref{eq:param} and different values of the SI breaking parameter $ \gamma $. Each dot corresponds to a separate run; the numerical uncertainties were always smaller than the point sizes. The dashed lines show power law fits with exponents as given in~\eqref{eq:non_SI_scaling}, \eqref{eq:p_theta_scaling} and~\eqref{eq:volume_scaling}, which are always approached as $ V \to \infty $. Whenever the scaling is $ \gamma $ independent, there are several data points which lie on top of each other.}
	\label{fig:non_SI}
\end{figure}

Figure~\ref{fig:non_SI} shows the numerical results for different values of $ \gamma > 0 $. Again, small $ \hat{R} $ and large $ V $ are generically realized for $ \delta \Phi \to 0 $, i.e.~if $ \Phi_\mathrm{GGP}$ is tuned close to $ \N $.
Evidently, both quantities again show a power law dependence on $ \delta\Phi $, with exponents which now depend on $ \gamma $. Empirically, we find the following laws,
\begin{equation} \label{eq:non_SI_scaling}
\hat{R} \propto 
\begin{cases}
\delta\Phi^{1 + 1/\gamma} \\
\delta\Phi^{2} 
\end{cases} , \qquad
V \propto 
\begin{cases}
\delta\Phi^{-1/\gamma} & \qquad (\text{for}\; 0< \gamma <1) \\
\delta\Phi^{-1}  & \qquad (\text{for}\; 1 < \gamma) 
\end{cases} ,
\end{equation}
as $ \delta\Phi \to 0 $. These are plotted in Figs.~\ref{fig:R} and~\ref{fig:V} as dashed lines, and evidently provide very good fits to the numerical data points. Note that the scalings for $ \gamma > 1 $ are the same as the ones obtained in the SI case $ \tau = 0 $. The transition to this generic scaling law occurs because for $ \gamma > 1 $ the finite width effects (which are independent of $ \gamma $) dominate, cf.~Sec.~\ref{sec:delta_limit}. Also note that combining the scaling relations for $ \hat{R} $ and $ V $ exactly reproduces the analytic prediction~\eqref{eq:R_scaling}.
For completeness, let us mention that the corresponding numerical coefficients for $ N_1 $ in~\eqref{eq:N12}, i.e.~the ratios $ N_1 / (\gamma\tau) $, were found in the range $ \sim 2 $ to $ 6 $.
Likewise, the scaling relations~\eqref{eq:p_theta_scaling} for $ p_\theta $, which are drawn as dashed lines in Fig.~\ref{fig:pTheta}, again agree very well with the data. Finally, Fig.~\ref{fig:phi} shows the relation between the dilaton evaluated at the brane and the volume, confirming~\eqref{eq:volume_scaling}.
%

With these results, we can now turn to the tuning question. For $ \gamma > 1 $, the discussion is exactly the same as for the SI case ($ \tau = 0 $) above, because the scaling relations are the same. But for $ \gamma < 1 $ there is a modification: Using the scaling relations~\eqref{eq:non_SI_scaling}, the phenomenological bound~\eqref{eq:RV_phen} now implies
\begin{equation}
10^{-120} \sim \delta\Phi^{1+2/\gamma} \,.
\end{equation} 
For $ \gamma \lesssim 1 $, $ \delta\Phi $ still has to be tuned tremendously close to zero; but for $ \gamma \ll 1 $, this is not the case anymore. Specifically, if we choose $ \gamma \approx 1/60 $ (which is not hierarchically small), this relation is already fulfilled if $ \delta\Phi \sim 0.1 $, i.e.~without any fine-tuning of model parameters. So we find the remarkable result that the near-SI tension is capable of producing a small 4D curvature and a large volume (as compared to the fundamental bulk scale) without fine-tuning, although this was not possible for a SI tension ($ \tau = 0 $). At first sight, this looks very promising. However, on closer inspection, there is an even bigger problem with the volume bound~\eqref{eq:V_phen} than before, since $ \gamma \sim 1/60 $ and $ \delta\Phi \sim 0.1 $ now yields $ V \sim 10^{60} $, exceeding the bound by 32 orders of magnitude. In turn, if we chose $ \gamma \sim 1/28 $, so that the volume satisfies the bound for $ \delta\Phi \sim 0.1 $, then $ \hat{R} \sim 10^{-57} \Mp^2 $, which is $ 63 $ orders of magnitude larger than its observational bound.

In summary, while it is possible to get small $ \hat{R} $ and large $ V $ without tuning $ \Phi_\mathrm{GGP} $ extremely close to $ \N $, it is not possible for both of them to satisfy their phenomenological bounds, in accordance with the general discussion in Sec.~\ref{sec:phen}.

Let us note that this possibility of getting a large volume without large parameter hierarchies was also recently observed in~\cite{Burgess:2015lda}, where the same model was studied in a dimensionally reduced, effective 4D theory. However, there it was also assumed that it would at the same time be possible to have $ \hat{R} $ within its bounds (possibly via some independent fine-tuning), so that the model could in this way at least address the electroweak hierarchy problem (albeit not the CC problem). Here we found that this is not possible, because $ \hat{R} $ and $ V $ are not independent, and so one cannot tune $ \hat{R} $ without at the same time ruining the value of $ V $. 

\section{Conclusion} \label{sec:concl}

The main result of our preceding work~\cite{Niedermann:2015via} was that the SLED model (with delta branes) only guarantees the existence of 4D flat solutions if the brane couplings respect the SI of the bulk theory, and that this comes at the price of a fine-tuning (or runaway), as expected~\cite{Weinberg:1988cp}. Here, we took one step further and asked how large the 4D curvature $ \hat{R} $ is for SI breaking couplings and the (more realistic) case of a finite brane width not below the fundamental 6D Planck length.


Specifically, we worked with a regularization which replaces the delta brane by a ring of stabilized circumference $ \ell $, and considered a SI breaking tension term parametrized as $ \mathcal{T}_+ = \lambda + \tau\, \re^{\gamma\phi_+} $. This type of dilaton-brane coupling is particularly interesting with respect to the CC problem as it allows to be close to SI without assuming an unnaturally small coefficient $\tau$. We then followed two complementary routes:

First, we analytically derived a formula for $ \hat{R} $. Motivated by the GGP solution, the extra space volume was then assumed to be proportional to $ \re^{-\phi_+} $. This resulted in the rigid relation~\eqref{eq:R_scaling} between $\hat R$ and the extra space volume $V$, consisting of two $ V $-dependent contributions to $ \hat{R} $ with unknown numerical constants of proportionality $ N_1 $ and $ N_2 $. They originate from the SI breaking dilaton coupling and the finite brane width, respectively.
Provided that $ N_{1/2} \sim 1 $, we found that either $ \hat{R} $ or $ V $ exceeds its phenomenological bound (by 36 or 12 orders of magnitude, respectively).

Second, we solved the full bulk-brane field equations numerically. By enforcing the correct boundary conditions at both branes, we were able to calculate all observables, in particular $ \hat{R} $ and $ V $, for given model parameters.
We thereby confirmed the analytically derived scaling relations without relying on any approximations and were able to explicitly compute the coefficients $ N_{1/2} $, indeed affirming $ N_{1/2} \sim 1 $.
The only way to get $ N_1 \ll 1 $ would be to either require SI brane couplings---which would ruin solar system tests due to a fifth force~\cite{Burgess:2015lda}---or to fine-tune (either $ \tau $ or $ \lambda $).
As for $ N_2 $, the only caveat is provided by allowing the brane width $ \ell $ to be much ($ \sim 18 $ orders of magnitude) smaller than the bulk Planck scale. This, however, would confront us with the problem how such a hierarchy could arise naturally, and whether one would have to take quantum gravity effects into account.


Moreover, the numerical analysis admitted an extensive discussion of the tuning issue. To be precise, we calculated the amount of tuning necessary to realize a large hierarchy between the bulk scale and $ V $, as is phenomenologically required according to \eqref{eq:V_phen}, with the following results:

\begin{itemize}
	\item For SI couplings ($\tau=0$) a sufficiently large $V$ is only achieved by tuning the total flux (or, equivalently, the brane tension) close to the corresponding GGP value with a precision of $ \sim 10^{-28} $. 
	
	\item If SI is broken explicitly by a $\phi$-dependent tension, it turns out that the tuning problem can in fact be avoided for near SI tension couplings $ \gamma \ll 1 $, in agreement with~\cite{Burgess:2015lda}. However, the phenomenological problem still persists (and even gets worse). Explicitly, for $ \gamma \sim 1/28 $, which yields the required volume without tuning, $ \hat{R} $ would be 63 orders of magnitude above its measured value.
\end{itemize}

In summary, there are no phenomenologically viable solutions in the SLED model if the brane width is not smaller than the fundamental bulk Planck length. But even if this were allowed, the required SI breaking dilaton coupling of the brane fields would always lead to a way too large 4D curvature or extra space volume, unless some sort of fine-tuning is at work.


\acknowledgments
We thank Cliff Burgess, Ross Diener, Stefan Hofmann, Tehseen Rug and Matthew Williams for many helpful discussions.
The work of FN was supported by TRR 33 ``The Dark Universe''.
The work of FN and RS was supported by the DFG cluster of excellence ``Origin and Structure of the Universe''.


\appendix

\section{Validity of Delta-Analysis}
\label{ap:Cliff}
The authors of~\cite{Burgess:2015kda} critically assessed our preceding work~\cite{Niedermann:2015via} based on a delta-analysis.\footnote{They only considered the case without BLF, so we will do the same here.} Specifically, they argued that the unregularized approach did not take into account a hidden metric dependence of the delta-function of the form
\begin{align}
\frac{\partial \delta^{(2)}(y)}{\partial g_{\theta\theta}}  =: C \,\frac{\delta^{(2)}(y)}{B_+^2} \;,
\end{align}
which would introduce an additional (localized) term in the $(\theta\theta)$-Einstein equation. In that case, 
the constant $C$ would be constrained by the radial Einstein equation \eqref{eq:einstein_rho} in terms of the brane tension; specifically, we find\footnote{This indeed agrees with the finding in~\cite{Burgess:2015kda} up to an irrelevant factor $-2$, which we think got somehow lost in~\cite{Burgess:2015kda}.}
\begin{align}\label{eq:C2}
\mathcal{T}_+ C \simeq - \frac{\kappa^2}{8\pi} \frac{\mathcal{T}'^2_+}{\left(1-\frac{\kappa^2 \mathcal{T}_+}{2 \pi}\right)}\;,
\end{align}
where higher order terms in $ \mathcal{T}'_+ $ were neglected.

The first important observation is that $ C $ vanishes for $ \mathcal{T}'_+ = 0 $. This shows that the concerns of~\cite{Burgess:2015kda} do not apply to the SI case. So one of the central results of~\cite{Niedermann:2015via}, namely that $ \hat{R} = 0 $ for SI delta branes (and not for dilaton-independent couplings, as had been claimed previously~\cite{Burgess:2011mt, Burgess:2011va}), is insensitive to this issue.

But it also looks as if assuming $C=0$, as implicitly done in~\cite{Niedermann:2015via}, would be in conflict with the SI breaking case $ \mathcal{T}'_+ \neq 0 $. This was exactly the argument given in~\cite{Burgess:2015kda}. However, there is a loophole to that reasoning: the right hand side of \eqref{eq:C2} depends on $\phi$ evaluated at the position of the delta brane, so we cannot make any final statement without knowing its value. In particular, $\phi_+$ could be such that the right hand side vanishes in the case of an infinitely thin brane.

The intuitive explanation for $ C \neq 0 $ in~\cite{Burgess:2015kda} was that a delta function should depend on the proper distance from the brane and thus implicitly on the off-brane metric. However, this picture is misleading since $ C $ is in fact not $ \partial\delta(y) / \partial g_{\rho\rho} $ (which vanishes!), but $ \partial\delta(y) / \partial g_{\theta\theta} $. Hence, in the parlance of~\cite{Burgess:2015kda} $ C $ corresponds to the delta function's knowledge about the azimuthal distance around a point. Equivalently, and more physically speaking, it is the azimuthal pressure of the point source.
This is obvious after noticing that the introduction of $ C $ is formally equivalent to introducing $ p_\theta $ as we did in our ring-regularization, upon identifying $ \lim\limits_{\epsilon\to0} p_\theta \equiv -2 \mathcal{T}_+(\phi) C $.
Either way, $ C \neq 0 $ seems to be rather unphysical.

While the analysis of \cite{Niedermann:2015via} is in line with the physical (but indeed more qualitative) argument that there is no well-defined notion of an angular pressure for an infinitely thin object, we think that a rigorous statement requires an explicit calculation of the right side of \eqref{eq:C2}. Since $ \phi $ can generically diverge at the non SI delta brane, this can only be done by first introducing a regularization of (dimensionless) width $ \epsilon $ and then letting $ \epsilon \to 0 $. This was (admittedly) not done in~\cite{Niedermann:2015via}, but neither in~\cite{Burgess:2015kda, Burgess:2015nka, Burgess:2015gba, Burgess:2015lda}. But it was done in this work, and we were able to give an unambiguous answer: For the relevant case of an exponential dilaton coupling,\footnote{Note that we checked this not only for the exponential tension coupling as discussed in the main text, but also for the analogous exponential BLF coupling.} $p_\theta \to 0$ in the delta limit (and thus $ C=0 $)---in accordance with our physical expectation. As a result, the old delta analysis correctly captures the physics of an exponential dilaton coupling.

However, it should be noted that whenever $ p_\theta \to 0 $, also $ \hat{R} \to 0 $, cf.~\eqref{eq:R_hat_delta_2} and~\eqref{eq:ptheta1}. As already mentioned in Sec.~\ref{sec:constraint}, this was not realized in the delta-analysis~\cite{Niedermann:2015via}, where it would have translated to the impossibility of breaking SI on a delta brane. But this would only have given yet another reason for studying the (more realistic) regularized setup, as we now did. Nonetheless, it is true that the delta formula for $ \hat{R} $ gives the correct leading nonzero contributions that arise for a regularized, near SI brane, as discussed in Sec.~\ref{sec:delta_limit}.

Now, let us be more specific and explicitly evaluate \eqref{eq:C2}. First, for all couplings studied, we verified numerically\footnote{Recall that, since $ \epsilon \equiv \ell^2 / V $, one way of realizing the delta limit is to take $ V \to \infty $.}
\begin{align} \label{eq:phi_plus}
\phi_+ \to - \infty \qquad (\text{for} \quad V \to \infty)\;.
\end{align}
We start with the physically relevant exponential coupling \eqref{eq:BLFcoupling} (as already discussed, this allows to be close to SI without tuning the coefficient). Then, Eq.~\eqref{eq:C2} implies a vanishing $C$ in the limit \eqref{eq:phi_plus}, hence proving that the loophole is realized. 

We also considered monomial couplings; physically, they are less interesting as they either lead to a diverging negative or super-critical tension in the limit \eqref{eq:phi_plus}. Nevertheless, even in these cases, we find $C \to 0$. For concreteness, consider a linear coupling in $\phi$: In that case, it is easy to check that the denominator in \eqref{eq:C2} diverges while the numerator is a constant, hence implying $C \to 0$ (albeit $ p_\theta \to \text{const} \neq 0 $, which we interpret as being caused by the pathological tension).

Of course, we could not check the validity of \eqref{eq:phi_plus} for  all possible couplings and there might very well be more complicated `designed potentials' with a different behavior. However, based on our previous findings we conjecture that these potentials either lead to a vanishing $C$ or again introduce some sort of pathology.

In summary, we agree with the formulas in \cite{Burgess:2015kda}, yet we come to a different conclusion based on a simple loophole that applies for both exponential and linear couplings (and probably for a much broader class which was beyond the scope of the present work). Let us stress that rigorously proving this result required to solve the full bulk-brane system. In particular, to show the validity of \eqref{eq:phi_plus}, it would not suffice to consider only a single brane without demanding the second brane to be physically well-defined.

Finally, it should be emphasized that we do agree---as discussed in great detail in this work---that $ p_\theta $ must be included for a brane of finite width, and has important consequences for the 4D curvature. Since this is the physically more relevant case anyhow, the delta-limit question becomes somewhat irrelevant. Still, the important achievement of~\cite{Niedermann:2015via}, namely the first correct identification of those BLF couplings which lead to $ \hat{R} = 0 $ (and the worries it raises), remains unaffected.

\bibliographystyle{utphys}
\bibliography{SLED_BLF_II}

\end{document}